\documentclass[prd, aps, nofootinbib, preprint, letterpaper]{revtex4-2}

\usepackage{mathtools}
\usepackage{color}
\definecolor{darkred}{RGB}{196,0,0}

\usepackage{amsmath, amssymb, amsfonts}
\usepackage[hidelinks]{hyperref}
\usepackage{ulem}
\usepackage{setspace}
\usepackage{extarrows}
    
\newcommand{\be}{\begin{equation}}
\newcommand{\ee}{\end{equation}}
\newcommand{\ba}{\begin{eqnarray}}
\newcommand{\ea}{\end{eqnarray}}

\newcommand{\la}{\label}

\newcommand{\wh}{{\hat\omega}}

\begin{document}

\title {Revisiting the collisional energy loss of a heavy quark in an anisotropic quark-gluon plasma}

\author{Mingda Cai$^{a,b}$, Yun Guo$^{a,b,*}$, and Enke Wang$^{c,d,\dagger}$}
\affiliation{
$^a$ Guangxi Key Laboratory of Nuclear Physics and Technology, Guangxi Normal University, Guilin 541004, China\\
$^b$ Department of Physics, Guangxi Normal University, Guilin, 541004, China\\
$^c$ State Key Laboratory of Nuclear Physics and Technology, Institute of Quantum Matter,
South China Normal University, Guangzhou 510006, China\\ 
$d$ Guangdong Basic Research Center of Excellence for Structure and Fundamental Interactions of Matter, Guangdong Provincial Key Laboratory of Nuclear Science, Guangzhou 510006, China
}

\renewcommand{\thefootnote}{\fnsymbol{footnote}}
\footnotetext[1]{yunguo@mailbox.gxnu.edu.cn}
\footnotetext[2]{wangek@scnu.edu.cn}
\renewcommand{\thefootnote}{\arabic{footnote}}

\begin{abstract}
We revisit the collisional energy loss of a heavy quark in a quark–gluon plasma with a small momentum-space anisotropy. Within the Braaten–Thoma framework, we obtain the hard contribution in closed form and reduce the soft contribution to one-dimensional integrals. Our result demonstrates analytically that, at linear order in the anisotropy, the dependence on the artificial separation scale cancels between the two contributions. By considering the unnormalized anisotropic distribution together with two variants that hold either the parton number density or the energy density at its isotropic value, we find that the angle-dependent part of the anisotropy correction is identical in all three cases and is proportional to the quadrupole $P_2(\cos\alpha)$, where $\alpha$ is the angle between the heavy-quark velocity and the anisotropy direction. In addition, the sign of the anisotropy correction, which determines whether the energy loss is enhanced or suppressed relative to the isotropic result, depends jointly on the quark speed, coupling, and propagation angle. In particular, at fixed energy density, only the quadrupole term survives. For a positive coefficient of $P_2(\cos\alpha)$, the correction enhances the energy loss for $\cos^2\alpha>1/3$ and suppresses it for $\cos^2\alpha<1/3$, with the opposite behavior for a negative coefficient.
\end{abstract}

\maketitle
\newpage

\section{Introduction}\la{intro}

Energetic partons produced in the initial hard scatterings of ultrarelativistic heavy-ion collisions can probe the hot and dense quark-gluon plasma (QGP) through their subsequent interactions with its constituents. The resulting modification of jets and high-transverse-momentum hadron spectra, which is commonly referred to as jet quenching, offers quantitative insight into the microscopic dynamics and transport properties of the QGP~\cite{Qin:2015srf,Baier:2000mf,Gyulassy:2003mc,Majumder:2010qh,Connors:2017ptx,Casalderrey-Solana:2007knd,Cao:2020wlm,Mehtar-Tani:2013pia}. Although medium-induced gluon radiation is essential for parton energy loss at sufficiently high energies, the suppression of small-angle radiation by the heavy-quark mass, known as the dead-cone effect, enhances the relative importance of elastic interactions~\cite{Dokshitzer:2001zm,Wicks:2005gt,Mustafa:2004dr,Zhang:2003wk}. Collisional energy loss can therefore compete with radiative energy loss over a substantial kinematic range and is important for an accurate description of the heavy-flavor propagation and relaxation in the QGP~\cite{Qin:2007rn,Schenke:2009gb}.

Extensive theoretical work has already been devoted to collisional energy loss. Thoma and Gyulassy investigated quark energy loss using the medium response and demonstrated the importance of screening in regulating infrared-sensitive interactions~\cite{Thoma:1990fm}. Braaten and Thoma subsequently developed a systematic calculation at leading order in the coupling~\cite{Braaten:1991jj,Braaten:1991we}.  By introducing an intermediate scale $q^*$~\cite{Braaten:1991dd}, the soft and hard momentum transfers can be treated separately. Soft interactions require a hard-thermal-loop (HTL) resummed propagator, whereas hard scatterings are evaluated with bare propagators. Peigné and Peshier later revisited the high-energy regime, identifying an additional collinear logarithm and incorporating running-coupling effects~\cite{Peigne:2007sd,Peigne:2008nd}. These foundational calculations assume an isotropic medium in thermal equilibrium, whose constituent quarks and gluons are described by equilibrium distribution functions.

However, the QGP produced in heavy-ion collisions need not remain locally isotropic throughout its evolution. Rapid longitudinal expansion can generate unequal longitudinal and transverse pressures, particularly during the early stages. An anisotropic parton distribution thus provides a more realistic description of these early stages than an equilibrium distribution. A widely used parametrization, introduced by Romatschke and Strickland, is obtained by stretching or squeezing an isotropic distribution along a preferred direction denoted by the unit vector ${\bf n}$~\cite{Romatschke:2003ms}. Within the HTL approximation, they analyzed the stable and unstable collective modes appearing in the anisotropic plasma~\cite{Romatschke:2003ms,Romatschke:2004jh}. Anisotropic distributions also serve as the starting point for the hydrodynamic formulations~\cite{Martinez:2010sc,Florkowski:2010cf}, which describe expansion around a locally anisotropic background. Collisional energy loss in an anisotropic plasma was first studied in Refs.~\cite{Romatschke:2003vc,Romatschke:2004au}, where the authors demonstrated a pronounced dependence of the energy loss on the propagation direction and discussed the energy gain associated with plasma instabilities. These studies establish that momentum-space anisotropy affects both the collective response of the medium and the energy exchange experienced by a heavy probe.

In this work, we revisit heavy-quark collisional energy loss in an anisotropic QGP within the widely used Braaten–Thoma framework, working at weak coupling and to first order in $\xi$, an adjustable parameter that quantifies the degree of momentum-space anisotropy. A central question is whether the cancellation of the artificial separation scale $q^*$ survives when the equilibrium distributions are replaced by anisotropic ones. We also examine the finite anisotropy correction as a function of the heavy-quark velocity, the coupling, and the incidence angle with respect to the anisotropy direction ${\bf n}$, which determines the directions of maximal/minimal energy loss and the conditions under which anisotropy enhances or suppresses the isotropic result. To distinguish changes in number/energy density from changes in momentum-space shape, we further compare the deformed distribution with the normalized ones in which either the parton number density or the energy density is held fixed at its thermal-equilibrium value. We point out that our emphasis differs from that of Refs.~\cite{Romatschke:2003vc,Romatschke:2004au}. Although those studies also adopted the soft-hard separation, they retain integral expressions without imposing the asymptotic scale expansions, thereby enabling numerical investigations at finite coupling. The resulting residual separation-scale dependence is treated using the principle of minimum sensitivity and scale variation to estimate uncertainty. Here, we focus on explicit analytical matching in the weak-coupling limit and on the angular dependence and overall sign of the leading anisotropy correction.

The rest of this paper is organized as follows. Sec.~\ref{defel} introduces the theoretical framework and derives the hard and soft contributions to the collisional energy loss in the small-anisotropy approximation. By combining these contributions, Sec.~\ref{numeric} demonstrates the cancellation of the separation scale, and examines the dependence of the energy loss on the heavy-quark velocity, coupling, and incident angle. The corresponding analyses at fixed parton number/energy density are presented in Sec.~\ref{fixed}. Sec.~\ref{summary} summarizes our findings, and some details of the hard scattering calculation are provided in App.~\ref{appa}.

\section{The theoretical framework to computing the collisional energy loss}\la{defel}

The rate of energy loss per unit distance for an energetic parton traversing the QGP is given by
\be\label{def}
-\Big(\frac{d E}{d x}\Big)=\frac{1}{v}\int_M^\infty
dE^\prime (E-E') \frac{d \Gamma}{d E'}\, .
\ee
The interaction rate $\Gamma(E)$ can be expressed in terms of the squared matrix element for the corresponding scattering processes $|\mathcal{M}|^2$ between the incident heavy quark with mass $M$ and the massless constituent particles in the hot medium, 
\ba
\Gamma (E)&=&\frac{1}{2E} \int \frac{d^3 {\bf{p}}'}{(2\pi)^3 2{E}'}\int \frac{d^3 {\bf k}}{(2\pi)^3 2k} f_{B/F}({\bf k})\int \frac{d^3 {\bf {k}}'}{(2\pi)^3 2{k}'}[1 \pm f_{B/F}({\bf k}')]\,\nonumber \\
&& \hspace{2cm}\times (2\pi)^4\delta^4(P+K-{P}'-{K}')\,\bigg(\frac{1}{6}\sum|\mathcal{M}|^2\bigg)\, ,
\label{interaction_rate}
\ea
where we have weighted $|\mathcal{M}|^2$ with the distribution functions of the medium partons, which include the Pauli-blocking or Bose-enhancement factor. In principle, these distribution functions are arbitrary which can be either in equilibrium or out of equilibrium. In addition, $P=(E,{\bf p})$ denotes the four-momentum of the incoming heavy quark, which scatters off a constituent quark or gluon with four-momentum $K=(k, {\bf k})$. Accordingly, the primed variables correspond to the energies or momenta of the outgoing heavy-quark and medium partons. In Eq.~(\ref{def}), $v\equiv|{\bf v}|=|{\bf p}|/E$ is the velocity of the incident heavy quark. In Eq.~(\ref{interaction_rate}), the sum runs over colors and spins of the incoming and outgoing partons.

\begin{figure}[htbp]
\begin{center}
\includegraphics[width=0.8\linewidth]{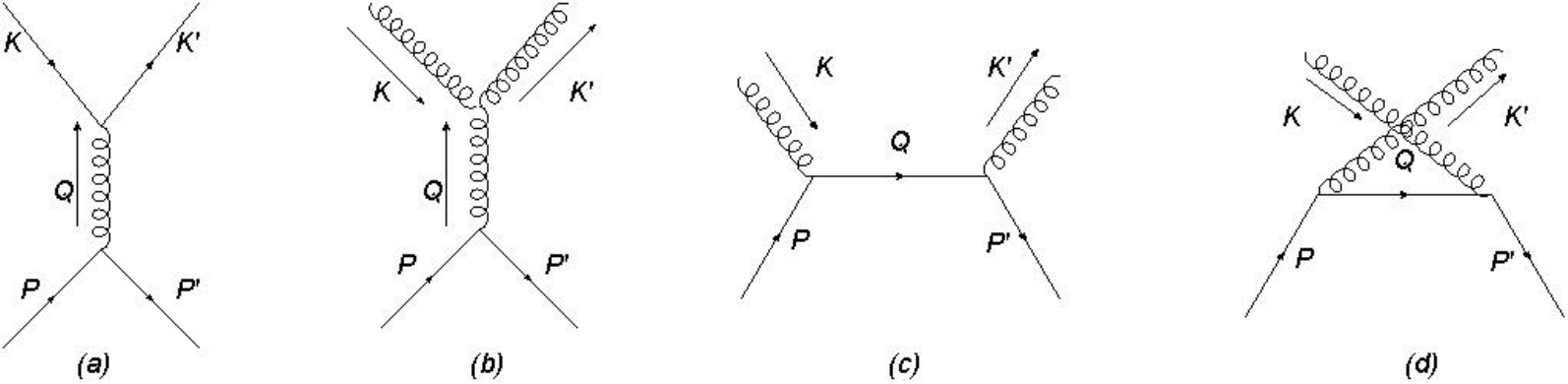}
\caption{The tree-level Feynman diagrams for the elastic scatterings $Q q \rightarrow Q q$ and $Q g\rightarrow Q g$.}
\label{lod}
\end{center}
\end{figure}

When computing the collisional energy loss, one encounters an infrared divergence arising from the $t$-channel diagrams shown in Figs.~\ref{lod}(a) and \ref{lod}(b). A standard way to remove this divergence is to impose a cutoff scale $q^*$ on the momentum transfer, thereby separating hard and soft scatterings\footnote{There also exit some other approaches to dealing with the infrared divergence; see, for example, Refs.~\cite{Djordjevic:2006tw,Guo:2024mgh,Cai:2025ntf,Grishmanovskii:2025mnc,Faraday:2024gzx,Carignano:2021mrn}.}. The hard contribution to the energy loss is obtained from a diagrammatic calculation using bare propagators, while infrared divergence in the soft scatterings can be regulated by employing the hard-thermal-loop (HTL) resummed gluon propagator. A key theoretical result is that, in the weak-coupling limit, the $q^*$ dependence is exactly canceled between the hard and soft contributions, so the total energy loss does not depend on this artificial cutoff~\cite{Braaten:1991jj,Braaten:1991we}. However, while the proof of the cancellation was carried out using equilibrium distribution functions, i.e., Fermi-Dirac distribution for quarks and Bose-Einstein distribution for gluons, it remains an open and theoretically interesting question whether the same cancellation can happen under non-equilibrium conditions.

In this work, we consider the following non-equilibrium distribution function~\cite{Romatschke:2003ms}
\be\label{anisodis}
f_{\rm aniso} ({\bf k})\equiv f_{\rm iso}\!\left(\frac{1}{\lambda}\sqrt{{\bf k}^2+\xi ({\bf k}\cdot {\bf n})^2}\right)\,.
\ee
which takes into account the rapid longitudinal expansion of the QGP, and thus exhibits an anisotropy in the momentum space. In Eq.~(\ref{anisodis}), $f_{\rm aniso} ({\bf k})$ can be obtained by stretching or squeezing an arbitrary isotropic distribution function $f_{\rm iso}$ along the direction of anisotropy denoted by the unit vector $\bf{n}$. Specifically, we consider $f_{\rm iso}$ to be the Fermi-Dirac or Bose-Einstein distribution function and assume $\bf{n}$ is parallel to the beam-line direction. Therefore, a contraction of the isotropic distribution in $\bf{n}$ direction is relevant to high-energy heavy-ion experiments. In this case, the adjustable parameter $\xi$ used to quantify the degree of momentum-space anisotropy is positive. In addition, $\lambda$ is a temperature-like scale that should be identified with the temperature $T$ only in the thermal equilibrium limit, however, we will continue to refer to $\lambda$ as temperature $T$ for simplicity.

Employing the small-$\xi$ approximation, the collisional energy loss can be formally expressed as
\be\label{elformal}
-\Big(\frac{dE}{dx}\Big)\equiv -\Big(\frac{dE}{dx}\Big)_{(0)}-\Big(\frac{dE}{dx}\Big)_{(1)}\, ,
\ee
where the first term on the right-hand side denotes the isotropic contribution, in which the distribution functions are the Fermi-Dirac and Bose-Einstein distributions, and the second term presents the anisotropy correction. 

\subsection{The hard contribution to the collisional energy loss}\la{hard}

To obtain the hard contribution to the collisional energy loss in an anisotropic plasma, one only needs to replace the isotropic/equilibrium distribution functions with anisotropic ones, since the squared matrix element evaluated with bare propagators remains the same as that in an equilibrium QGP. However, the introduction of a preferred anisotropy direction makes the calculation less straightforward than one might expect from the experience gained in the equilibrium case. In particular, the collisional energy loss is no longer isotropic, and the previously used trick of averaging over all possible directions of the incident velocity $\mathbf{v}$ no longer works in the anisotropic case~\cite{Braaten:1991jj}. On the other hand, considering the small-$\xi$ approximation, the distribution function in Eq.~(\ref{anisodis}) can be expanded up to the linear order in $\xi$, yielding
\be\label{aniso-ex}
f_{\rm aniso} ({\bf k})\approx
 f_{\rm iso}(k)-\xi\frac{({\bf n}\cdot{\bf k})^2}{2 k \lambda}(1\pm f_{\rm iso}(k))f_{\rm iso}(k) 
\equiv f_{(0)}(k)+f_{(1)}({\bf k})\,.
\ee
For the Bose–Einstein distribution, we use the ``$+$" sign and set $f_{(0)}(k)=n_{B}(k)$ in the above expansion. Correspondingly, the ``$-$" sign applies to the Fermi–Dirac distribution, with $f_{(0)}(k)=n_{F}(k)$. Accordingly, we expand the thermal weight factor $f_{B/F}({\bf k}) [1 \pm f_{B/F}({\bf k}')]$ in Eq.~(\ref{interaction_rate}) to the linear order in $\xi$, yielding the following expression
\be\label{twex}
f_{B/F}({\bf k}) [1 \pm f_{B/F}({\bf k}')]\approx f_{(0)}(k)+f_{(1)}({\bf k})\pm f_{(0)}(k)f_{(0)}(k')\pm f_{(1)}({\bf k})f_{(0)}(k')\pm f_{(0)}(k)f_{(1)}({\bf k}')\,.
\ee
The isotropic part in the collisional energy loss $-\big(d E/d x\big)_{\rm hard(0)}$ has been calculated previously under the approximation\footnote{Note that this approximation holds in the heavy quark limit, whereas in the ultra-relativistic regime ($E \gg M^2/T$) further corrections are required~\cite{Peigne:2008nd}.} that $T\ll M$, $T/E \ll v$ and $E\ll M^2/T$~\cite{Braaten:1991we}. It is important to note that given the squared matrix elements obtained in this approximation, terms quadratic in the distribution functions in Eq.~(\ref{twex}) do not contribute to the collisional energy loss, which significantly simplifies our calculation. This follows from the fact that contributions to the collisional energy loss involving those terms vanish after integrating over the momenta, owing to the symmetry under the interchange ${\bf k}\leftrightarrow {\bf k}'$.

For completeness, we list the corresponding isotropic result. The quark-quark scattering contribution reads
\be\label{loqq}
-\Big(\frac{dE}{dx}\Big) _ {{\rm hard} (0)} ^ {Q q}  =  \frac {g^ {4} T ^ {2}}{6 \pi} \frac{N_f}{6}\bigg [{\cal J}_0(v) \left( \ln \frac {T}{q ^ {*}}+{\rm C}_0^q\right) +{\cal F}_0(v)\bigg]\,,
\ee
where
\be
{\cal J}_0(v)= \frac {1}{v} - \frac {1 - v ^ {2}}{2v ^ {2}}\ln\frac{1+v}{1-v}\,,
\ee
and the constant ${\rm C}_0^q=\ln (4\pi)+12 \zeta^\prime(-1)\approx 0.546$ with $\zeta^\prime$ the derivative of the Riemann zeta function. The function ${\cal F}_0(v)$, which involves the dilogarithm $\mathrm{Li}_{2}(x)$, is defined as follows
\be
{\cal F}_0(v)=\frac{1}{6v^2}\big[3(1-v^2)f_1(v)+3v f_2(v)+v(3-4v^2)\big]\,,
\ee
with
\ba
f_1(v)&=& \frac{\pi^2}{12}+\mathrm{Li}_{2}\left(\frac{v+1}{v-1}\right)+\frac{1}{4}\left(\ln\frac{1+v}{1-v}-2\right)\ln\frac{1+v}{1-v}\,,\\ \nonumber
f_2(v)&=&2 \ln 2-\ln\left(1-v^2\right)\,.
\ea
In addition, the contribution from quark-gluon scattering has two sources. One is $t$-channel scattering, which can be computed similarly to the quark-quark scattering case. The other corresponds to $s$- and $u$-channel scattering as shown in Figs.~\ref{lod}(c) and \ref{lod}(d). Although these two channels are free of infrared divergence and therefore not relevant for demonstrating the cancellation of the momentum cutoff, they contribute non-negligibly to the collisional energy loss compared to the $t$-channel contribution. Summing all these contributions, the collisional energy loss from quark-gluon scattering takes a form similar to that of Eq.~(\ref{loqq}), up to an overall factor of $N_f/6$ and a shift in the constant ${\rm C}_0^q$. It can be expressed as
\be\label{lo}
-\Big(\frac{dE}{dx}\Big)^{Qg}_{{\rm hard} (0)}  =  \frac {g^ {4} T^2 }{6 \pi} \bigg [{\cal J}_0(v) \left( \ln \frac {T}{q ^ {*}}+{\rm C}_0^g\right) +{\cal F}_0(v)\bigg]\,,  
\ee
where ${\rm C}_0^g={\rm C}_0^q-\ln 2$. Notice that the interference between $t$- and $s$/$u$-channel is neglected under the above mentioned approximation, which will also be adopted in computing the collisional energy loss in an anisotropic QCD plasma. 

It is straightforward to separate the isotropic energy loss into an infrared-divergent $q^*$-dependent part and a finite part. The corresponding results are listed as follows
\be
-{\Big(\frac{dE}{dx}\Big)}_{\rm {hard (0)}}^{{\rm div.}} =  \frac{g^2  m_D^2 }{6\pi} {\cal J}_0(v) \ln\frac{T}{q^*}\, .
\label{divhard0}
\ee 
\be
-{\Big(\frac{dE}{dx}\Big)}_{\rm {hard (0)}}^{{\rm fin.}}= \frac{g^2 m_D^2  }{6\pi} \Big[{\cal J}_0(v){\rm C}_{\rm h} + {\cal F}_0(v)\Big]\, .
\label{finhard0}
\ee
where we define ${\rm C}_{\rm h}={\rm C}_0^q-\ln 2/(1+N_f/6)$ and the perturbative Debye mass is given by $m_D^2=g^2T^2(1+N_f/6)$.

We will focus on the second term on the right-hand side in Eq.~(\ref{elformal}). It is proportional to $\xi$, representing the small-anisotropy correction to the isotropic collisional energy loss. Explicitly, we have
\ba
-\Big(\frac{dE}{dx}\Big)_{{\rm hard}(1)}&=&-\frac{\xi}{\lambda v}\frac{\pi}{16 E^2} \int \frac{d^3 {\bf k}}{(2\pi)^3 } \frac{({\bf n}\cdot{\bf k})^2}{k^2}(1\pm f_{(0)}(k))f_{(0)}(k) \int \frac{d^3 {\bf {k}}'}{(2\pi)^3 } \frac{\omega}{k^\prime} \,\nonumber \\
&\times& \bigg(\frac{1}{6}\sum|\mathcal{M}|^2\bigg)\delta(\omega-{\bf v}\cdot{\bf q})\, ,
\label{defhard}
\ea
where the four-momentum transfer $Q=K^\prime-K=(\omega,{\bf q})$. To carry out the above integrals, we can use the fact that the energy loss depends only on the angle $\alpha$ between ${\bf v}$ and ${\bf n}$, therefore, we can  average over the solid angles of ${\bf v}$ and ${\bf n}$ while keeping $\alpha$ fixed. Consequently, the following identity can be used to simplify the calculation,
\be\label{id}
-{\Big(\frac{dE}{dx}\Big)}_{\rm {(1)}}=-\frac{1}{8 \pi^2}\int_{\Omega_{\bf v}}\int_{\Omega_{\bf n}}{\Big(\frac{dE}{dx}\Big)}_{\rm {(1)}} \delta\left(\frac{{\bf v}\cdot {\bf n}}{v}-\cos \alpha\right)\, .
\ee
Upon averaging over these angles, the integrals over ${\bf k}$ and ${\bf k}'$ no longer depend on the orientations of ${\bf v}$ and ${\bf n}$. Following the same procedure as in the isotropic case, we can then change the integration variables to $k$, $q$, and $\omega$ as follows
\ba\label{changevah}
\frac{1}{2(2\pi)^2}\int \frac{d^3 \mathbf{k}}{k} \int \frac{d^3\mathbf{k}'}{k^\prime} \theta(q-q^*) &\to& \int_{\frac{1+v}{2}{q^*}}^{\infty} d k\int_{ q^*}^{\frac{2k}{1+v}} q dq \int_{-v q}^{v q} d\omega 
+\int_{\frac{1+v}{2}{ q^*}}^{\infty} d k\int_{\frac{2k}{1+v}}^{\frac{2k}{1-v}}q dq \int_{q-2k}^{v q} d\omega 
\nonumber \\&+&\int_{\frac{1-v}{2}{q^*}}^{\frac{1+v}{2}{q^*}} d k\int_{q^*}^{\frac{2k}{1-v}} q dq \int_{q-2k}^{v q} d\omega\,.
\ea
To obtain an analytical result for the anisotropy correction, we further assume $q^* \ll T$ and set $q^* = 0$ in the integration limits, provided this does not lead to divergence. We thus obtain a simplified version of Eq.~(\ref{changevah}) which is given by
\be\label{changeva}
\frac{1}{2(2\pi)^2}\int \frac{d^3 \mathbf{k}}{k} \int \frac{d^3\mathbf{k}'}{k^\prime} \theta(q-q^*)  \to  \int_{0}^{\infty} dk \left( \int_{q^*}^{\frac{2k}{1+v}} q dq \int_{-vq}^{vq} d\omega 
+ \int_{\frac{2k}{1+v}}^{\frac{2k}{1-v}}q dq\int_{q-2k}^{vq} d\omega \right) 
\, 
\ee

Using the explicit form of the squared matrix element for the quark-quark scattering, the corresponding energy loss reads  
\ba
-{\Big(\frac{dE}{dx}\Big)}_{\rm {hard (1)}}^{Qq} &=& -\xi \frac{g^4  T^2 }{18\pi}\frac{N_f}{6} \bigg\{\Big[{\cal J}_0(v) - (1 - 3 \cos^2\alpha) {\cal J}_1(v)\Big]\bigg(\ln\frac{T}{q^*}+ {\rm C}_1^q \bigg)\nonumber \\
&+&{\cal F}_0(v)-(1 - 3 \cos^2\alpha){\cal F}_1(v)\bigg\}\, ,
\label{qqtEhard}
\ea
where 
\be
{\cal J}_1(v)=\frac{1-v^2}{2v^2} \left(\frac{3}{v}- \frac {3 - v ^ {2}}{2v ^ {2}}\ln\frac{1+v}{1-v}\right)
\ee
and the constant ${\rm C}_1^q$ is defined as ${\rm C}_1^q={\rm C}_0^q+1/2\approx 1.046$ and ${\cal F}_1(v)$ takes the following form
\be
{\cal F}_1(v)=\frac{1}{4v^4}\big[(3-4v^2+v^4)f_1(v)+3v(1-v^2)f_2(v)+v (3-5v^2+32v^4/15-16v^6/35)\big]\,.
\ee
In addition, the contributions from quark-gluon scattering are obtained as
\ba
-{\Big(\frac{dE}{dx}\Big)}_{\rm {hard (1)}}^{Qg(t)}&=& -\xi \frac{g^4  T^2 }{18\pi} \bigg\{\Big[{\cal J}_0(v) - (1-3 \cos^2\alpha) {\cal J}_1(v)\Big]\bigg(\ln\frac{T}{q^*}+ {\rm C}_1^{g(t)} \bigg)\nonumber \\
&+&{\cal F}_0(v)-(1 - 3 \cos^2\alpha){\cal F}_1(v)\bigg\}\, ,
\label{qgtEhard}
\ea
with $ {\rm C}_1^{g(t)}={\rm C}_1^q-1/2-\ln 2$, and
\be
-{\Big(\frac{dE}{dx}\Big)}_{\rm {hard (1)}}^{Qg(s,u)}= -\xi \frac{g^4  T^2 }{18\pi} \Big[{\cal J}_0(v) - (1-3 \cos^2\alpha) {\cal J}_1(v)\Big] {\rm C}_1^{g(s,u)}\, ,
\label{qgsuEhard}
\ee
with ${\rm C}_1^{g(s,u)}=1/2$. It is interesting to note that, starting from the quark-quark scattering contribution, exactly the same modifications yield the quark-gluon scattering contribution for both the isotropic energy loss $-{\big({dE}/{dx}\big)}_{\rm {(0)}}$ and the corresponding small-$\xi$ correction $-{\big({dE}/{dx}\big)}_{\rm {(1)}}$. Specifically, the overall factor $N_f/6$ in Eq.~(\ref{loqq}) or (\ref{qqtEhard}) is replaced by $1$. For $t$-channel quark-gluon scattering, the result is obtained by shifting the constants ${\rm C}_{0/1}^q$ by $-1/2-\ln 2$; for the $s$- and $u$-channel contributions, in addition to dropping both the logarithmic term and the function $\mathcal{F}_0(v)$ (in the isotropic result) or $\mathcal{F}_0(v)-(1-3\cos^2\alpha)\mathcal{F}_1(v)$ (in the
linear anisotropy correction), the constants ${\rm C}_{0/1}^q$ are changed to $1/2$. Combining Eqs.~(\ref{qgtEhard}) and (\ref{qgsuEhard}), we can express the small-$\xi$ correction to the isotropic energy loss induced by quark-gluon scattering as
\ba
-{\Big(\frac{dE}{dx}\Big)}_{\rm {hard(1)}}^{Qg}&=& -\xi \frac{g^4  T^2 }{18\pi} \bigg\{\Big[{\cal J}_0(v) - (1-3 \cos^2\alpha){\cal J}_1(v)\Big]\bigg(\ln\frac{T}{q^*}+ {\rm C}_1^{g} \bigg)\nonumber \\
&+&{\cal F}_0(v)-(1 - 3 \cos^2\alpha){\cal F}_1(v)\bigg\}\, ,
\label{qgEhard}
\ea
with ${\rm C}_1^{g}={\rm C}_1^{g(t)}+{\rm C}_1^{g(s,u)}={\rm C}_1^{q}-\ln 2$. Since these analytical results for $-{\big({dE}/{dx}\big)}_{\rm {(1)}}$ are entirely new, we provide the calculational details in App.~\ref{appa} for interested readers.

The divergent part that depends on $q^*$ in the small-$\xi$ correction is also proportional to $m_D^2$ which is given by
\be
-{\Big(\frac{dE}{dx}\Big)}_{\rm {hard(1)}}^{{\rm div.}}= -\xi \frac{g^2  m_D^2 }{18\pi} \Big[{\cal J}_0(v) - (1-3 \cos^2\alpha) {\cal J}_1(v)\Big]\ln\frac{T}{q^*}\, .
\label{divhard}
\ee
Accordingly, the finite part can be written as 
\ba
-{\Big(\frac{dE}{dx}\Big)}_{\rm {hard(1)}}^{{\rm fin.}} &=& -\xi \frac{g^2 m_D^2}{18\pi} \bigg\{\Big[{\cal J}_0(v) - (1-3 \cos^2\alpha) {\cal J}_1(v)\Big]\Big({\rm C}_{\rm h}+\frac{1}{2}\Big)\nonumber \\
&+&{\cal F}_0(v)-(1 - 3 \cos^2\alpha){\cal F}_1(v)\bigg\}\,.
\label{finhard}
\ea

\subsection{The soft contribution to the collisional energy loss}\la{soft}

The collisional energy loss from soft scattering can be computed in different ways. One approach is a field-theoretic calculation, as first carried out in Ref.~\cite{Braaten:1991jj}. Alternatively, one can employ a method borrowed from plasma physics~\cite{Thoma:1990fm,Romatschke:2004au}. In both cases, the HTL-resummed gluon propagator must be adopted to regulate the infrared divergence. It has been verified that different calculational approaches give identical results for the collisional energy loss~\cite{Thoma:2000dx}. In the presence of momentum-space anisotropy, it is convenient to use the latter method to perform the calculation. Therefore, we can start from the following definition of the collisional energy loss
\be\label{defsoft}
-\Big(\frac{dE}{dx}\Big)_{\rm soft} = \frac{4g^2}{3v} \, {\rm Im}\int\frac{d^3\mathbf{q}}{(2\pi)^3}( {\mathbf{v}}\cdot\mathbf{q})v^i\big[\Delta^{ij}(\xi,Q)
-{(\Delta_{\rm bare})}^{ij}(Q)\big]v^j \theta(q^*-q)\Big|_{\omega={\bf v}\cdot{\bf q}}\, ,
\ee
where $\Delta^{ij}(\xi,Q)$ is the HTL-resummed gluon propagator in the temporal axial gauge for an anisotropic QGP, while ${(\Delta_{\rm bare})}^{ij}(Q)$ denotes the bare gluon propagator in the same gauge. Within the small-$\xi$ approximation, the expansion of $\Delta^{ij}(\xi,Q)$ can formally expressed as
\be
\Delta^{ij}(\xi,Q)=\Delta_{(0)}^{ij}(Q)+\xi\Delta_{(1)}^{ij}(Q)+\mathcal O(\xi^{2})\, ,
\ee
where the leading-order term, i.e., the isotropic part reads
\be\label{diso}
\Delta_{(0)}^{ij}(Q)=\frac{1}{q^{2}-\omega^{2}+m_D^2\,\Pi_T(\wh)}A^{ij} +\frac{1}{ \wh^2\big(m_D^2 \Pi_L(\wh)-q^2\big)}B^{ij}\, ,
\ee
and the anisotropy correction to the isotropic gluon propagator is given by~\cite{Guo:2008da}
\ba\label{D1}
\Delta_{(1)}^{ij}(Q)&=&-m_D^2\bigg[\frac{ 1}{\big(q^{2}-\omega^{2}+m_D^2\,\Pi_T(\wh)\big)^2}\big(\alpha_{(1)}A^{ij}+\gamma_{(1)}C^{ij}\big)+\frac{\beta_{(1)}}{\wh^4\big(m_D^2 \Pi_L(\wh)-q^2\big)^2}B^{ij}\nonumber  \\ &+&\frac{\delta_{(1)}}{\wh^2\big(q^{2}-\omega^{2}+m_D^2\,\Pi_T(\wh)\big)\big(m_D^2 \Pi_L(\wh)-q^2\big)}D^{ij}\bigg]\, .
\ea
In the above equations, we define $\hat\omega=\omega/q$, and the transverse and longitudinal gluon self-energies can be written as
\ba
\Pi_T(\wh) &=& \frac{\wh^2}{2}\bigg(1-\frac{\wh^2-1}{2\wh}\ln\frac{\wh+1}{\wh-1}\bigg)\,,\nonumber \\
\Pi_L(\wh)& = & -1+\frac{\wh}{2}\ln\frac{\wh+1}{\wh-1}\, .
\label{se}
\ea
In addition, the four tensor bases are defined by
\be
A^{ij}=\delta^{ij}-\frac{q^i q^j}{q^2}\,,\qquad
B^{ij}=\frac{q^i q^j}{q^2}\,, \qquad
C^{ij}=\frac{\tilde n^i\tilde n^j}{\tilde{ n}^2}\, ,\qquad
D^{ij}=q^i\tilde n^j+q^j\tilde n^i\, ,
\ee
with
\be
\tilde n^i=A^{ij}n^j=n^i-\frac{q^i({\bf q}\cdot\mathbf n)}{q^2}\,,
\ee
and the associated structure functions take the following forms~\cite{Romatschke:2003ms}
\ba
\nonumber \alpha_{(1)}&=&\frac{\hat\omega^2}{12}(3+5\cos 2\theta_n)-\frac16(1+\cos 2\theta_n)+\frac{1}{4}\Pi_T(\hat\omega)\left[(1+3\cos 2\theta_n)-\hat\omega^2(3+5\cos 2\theta_n)\right]\,,\\ \nonumber
\beta_{(1)}&=&\hat\omega^2\left\{\frac16(1+3\cos 2\theta_n)\right.\left.+\Pi_L(\hat\omega)\left[\cos 2\theta_n-\frac{\hat\omega^2}{2}(1+3\cos 2\theta_n)\right]\right\}\,,\\ \nonumber
\gamma_{(1)}&=&\frac13\bigl(3\Pi_T(\hat\omega)-1\bigr)(\hat\omega^2-1)\sin^2\theta_n\,,\\ 
\delta_{(1)}&=&\frac{\cos\theta_n}{3q}\left[4\hat\omega^2+3\Pi_T(\hat\omega)(1-4\hat\omega^2)\right]\,.
\ea
Here, $\theta_n$ denotes the angle between $\mathbf n$ and $\mathbf q$.

The leading-order contribution to the collisional energy loss, corresponding to vanishing $\xi$, has been obtained previously~\cite{Braaten:1991we}. We also list the result for completeness. The divergent part, which is exactly cancelled by the corresponding hard contribution given by Eq.~(\ref{divhard0}), is found to be
\be
-{\Big(\frac{dE}{dx}\Big)}_{\rm {soft(0)}}^{{\rm div.}}= - \frac{g^2  m_D^2 }{6\pi} {\cal J}_0(v) \ln\frac{T}{q^*}\, .
\label{qgdivsoft0}
\ee 
Furthermore, to obtain the non-vanishing leading-order result in the small-coupling limit for an isotropic QGP, we should add the following finite contribution to Eq.~(\ref{finhard0}),
\be
-{\Big(\frac{dE}{dx}\Big)}_{\rm {soft(0)}}^{{\rm fin.}}=-\frac{g^2m_D^2}{6\pi}\Big[{\cal J}_0(v){\rm C}_{\rm s}(g)+{\cal G}_{0}(v)
\Big]\, ,
\label{qgfinsoft0}
\ee
where ${\rm C}_{\rm s}(g)=\ln \Big(g \sqrt{1+N_f/6}\Big)$ and
\be
{\cal G}_{0}(v)=\frac{1}{v^2} \int_0^v
\,d\wh\, \wh^2 \Big[\frac{v^2-\wh^2}{2(1-\wh^2)}\mathrm{R}_T(\wh)+\mathrm{R}_L(\wh)\Big]\,,
\ee
with
\ba
\mathrm{R}_T(\wh)&=&
\mathrm{Q}_T(\wh)\Theta_T(\wh)
+\ln\frac{\pi\wh}{4}
+\ln\sqrt{1+\mathrm{Q}_T^2(\wh)}\,, \\ \nonumber
\mathrm{R}_L(\wh)&=&
\mathrm{Q}_L(\wh)\Theta_L(\wh)
+\ln\frac{\pi\wh}{2}
+\ln\sqrt{1+\mathrm{Q}_L^2(\wh)}\,,
\ea
and
\ba
\mathrm{Q} _ {T} (\wh) &=& \frac{1}{\pi}\left(\ln \frac {1 +\wh}{1 - \wh} + \frac {2 \wh}{1 - \wh^ {2}}\right) \,,\qquad \Theta_T(\wh)=\frac{\pi}{2}-\arctan {\rm Q}_T(\wh)\, ,\\ \nonumber
\mathrm{Q} _ {L} (\wh) &=& \frac{1}{\pi}\left( - \ln \frac {1 + \wh}{1 -\wh}+\frac {2}{\wh}\right) \,, \qquad\quad
\Theta_L(\wh)=\frac{\pi}{2}-\arctan {\rm Q}_L(\wh)\,.
\ea

The calculation of the anisotropy correction to the above result turns out to be tedious. We therefore only outline the key steps. In fact, for both the isotropic energy loss and its small-$\xi$ correction, the integration over $q$ in Eq.~(\ref{defsoft}) can be carried out analytically, with the logarithmic divergence regulated by the cutoff $q^*$. The remaining integrals over the solid angle can be simplified by applying the identity in Eq.~(\ref{id}). Consequently, we need to evaluate the following multiple integrals
\be
\int d \Omega_{q} \int d \Omega_{v} \int d \Omega_{n}\, {\cal K}_{\rm int}(\cos \theta_v,\cos \theta_n)\delta\left(\frac{{\bf v}\cdot {\bf n}}{v}-\cos \alpha\right)\, ,
\ee
where the integrand, represented symbolically as ${\cal K}_{\rm int}$, is a complicated function of $\cos \theta_v$ and $\cos \theta_n$ with $\theta_v$ the angle between ${\bf v}$ and ${\bf q}$. Despite its overall complexity, ${\cal K}_{\rm int}$ depends on $\cos \theta_n$ only linearly and quadratically, one can first integrate over the solid angle of ${\bf n}$ using the integral formula,
\begin{equation}\la{int}
\int d \Omega_{n}\,(\cos \theta_n)^s\delta\left(\frac{{\bf v}\cdot {\bf n}}{v}-\cos \alpha\right)=\left\{
\begin{aligned}
&2\pi \, &{\rm for}\quad s=0\\
&2\pi \cos \theta_v \cos \alpha \, &{\rm for}\quad s=1\\
&\frac{\pi}{2}\Big[1+\cos^2 \theta_v-(1-3\cos^2 \theta_v) \cos 2\alpha\Big] &\,{\rm for}\quad s=2 \\
\end{aligned}
\right.\,.
\end{equation}
The resulting integrand depends only on $\cos \theta_v= \wh/v$. As a result, the double solid-angle integral $\int d \Omega_{q} \int d \Omega_{v}$ reduces to 
\be
8\pi^2 \int_{-1}^{1} d \cos \theta_v=\frac{8\pi^2}{v} \int_{-v}^{v} d \wh\, .
\ee
This indicates that the soft contribution to $-{\big({dE}/{dx}\big)}_{\rm {(1)}}$ can be expressed as a one-dimensional integral, just as in the isotropic case discussed above. The corresponding result is
\ba\label{soft1}
&&-{\Big(\frac{dE}{dx}\Big)}_{\rm {soft(1)}}=\xi \frac{g^2 m_D^2}{18 \pi^2 v^2}{\rm Im} \int_{-v}^v d \wh\, \wh \Bigg\{\frac{v^2-\hat{\omega}^2}{(1-\hat{\omega}^2)^2}
\ln\left(1+\frac{(q^*/m_D)^2(1-\hat{\omega}^2)}{\Pi_T(\wh)}\right)\big[
\Pi_T(\wh) \nonumber \\
&&\quad
\,
-\frac{1-3\cos^2\alpha}{12v^2}g_1(v,\wh)\big]+ \ln\left(1-\frac{(q^*/m_D)^2}{\Pi_L(\wh)}\right) \big[
\Pi_L(\wh)-\frac{1-3\cos^2\alpha}{6v^2}g_2(v,\wh)\big]
\nonumber \\ 
&&\quad-\,\Pi_L(\wh)\big[1-\frac{1-3\cos^2\alpha}{2 v^2}g_3(v,\wh)\big]-\frac{v^2-\hat{\omega}^2}
{ (1-\hat{\omega}^2)^2}\Pi_T(\wh)
\big[1-
\frac{1-3\cos^2\alpha}{4 v^2}g_4(v,\wh)\big]\Bigg\}\,,\nonumber 
\\
\ea
where we omit terms that vanish in the limit $q^*\gg m_D$ and define
\ba
g_1(v,\wh)&=&-(1-\hat{\omega}^2) \frac{4}{1-3\Pi_T(\wh)}+\big[4-6v^2+(8+9v^2)\hat{\omega}^2-21{\wh}^4\big]+3\Pi_T(\wh) \nonumber \\
&\times&\big[4+7v^2-(35+9v^2)\hat{\omega}^2+37\hat{\omega}^4\big]\,,
 \nonumber \\
 g_2(v,\wh)&=&-\frac{v^2-\wh^2}{1-\wh^2}\frac{4}{1+3(1-\wh^2)\Pi_L(\wh)}+\frac{7v^2-(13+3v^2)\hat{\omega}^2+9\hat{\omega}^4}{1-\wh^2}+3\Pi_L(\wh) \nonumber \\
&\times&\big[4v^2-(8+11v^2)\wh^2+17\wh^4\big]\,, \nonumber \\
g_3(v,\wh)&=&2v^2-(3v^2+6)\hat{\omega}^2+9\hat{\omega}^4\,, \nonumber \\
g_4(v,\wh)&=&7v^2-(15+9v^2)\hat{\omega}^2+21\hat{\omega}^4\,.
\ea

To explicitly demonstrate the cancellation of the momentum cutoff $q^*$, we also need to expand the logarithms in Eq.~(\ref{soft1}) in the small-coupling limit $q^* \gg gT$. Consequently, the logarithms can be approximated by
\ba
\ln \bigg[1-\frac{(q^*/m_D)^2}{\Pi_L(\wh)}\bigg]  &\approx& 2\ln (q^*/m_D) - \ln\left(-\Pi_L(\wh)\right)\, , \nonumber \\
\ln \bigg[1+\frac{(q^*/m_D)^2(1-\hat{\omega}^2)}{\Pi_T(\wh)}\bigg] &\approx& 2\ln (q^*/m_D) - \ln\Pi_T(\wh)+ \ln\left(1-\wh^2\right)\, .
\ea
According to Eq.~(\ref{soft1}), we need to take the imaginary part of the integral over $\wh$. To this end, it is useful to write the following expressions with the real and imaginary parts shown explicitly,
\ba
&&\Pi_T(\wh)=-\frac{\wh}{4}(\wh^2-1)\pi \big[{\rm Q}_T(\wh)-i\big]\, ,\qquad \Pi_L(\wh)=-\frac{\wh}{2}\pi \big[{\rm Q}_L(\wh)+i\big]\, ,\nonumber \\
&&  \ln\left(1-\wh^2\right)- \ln\Pi_T(\wh)= -\ln \frac{\pi |\wh|}{4}-\ln \sqrt{{\rm Q}_T^2(\wh)+1}+i \big[\pm \frac{\pi}{2}-\arctan{\rm Q}_T(\wh)\big]\, ,\nonumber \\
&&  \ln \left(-\Pi_L(\wh)\right)= \ln \frac{\pi |\wh|}{2}+\ln \sqrt{{\rm Q}_L^2(\wh)+1}+i \big[\pm \frac{\pi}{2}-\arctan{\rm Q}_L(\wh)\big]\,.
\ea
In the above equations, the $+$ sign applies when $\hat{\omega} > 0$, while the $-$ sign applies when $\hat{\omega} < 0$.

With the above discussions, it is straightforward to derive the anisotropy correction due to soft scatterings. For the $q^*$-dependent part, the integral over $\wh$ in Eq.~(\ref{soft1}) can be done analytically and the result reads
\be
-{\Big(\frac{dE}{dx}\Big)}_{\rm soft {(1)}}^{{\rm div.}}= \xi \frac{g^2  m_D^2 }{18\pi} \Big[{\cal J}_0(v) - (1-3 \cos^2\alpha) {\cal J}_1(v)\Big]\ln\frac{T}{q^*}\, .
\label{divsoft}
\ee
In addition, the finite part takes a form analogous to that of the corresponding hard contribution,
\ba
-{\Big(\frac{dE}{dx}\Big)}_{\rm soft {(1)}}^{{\rm fin.}}&=&\xi \frac{g^2 m_D^2}{18\pi}\bigg\{\Big[{\cal J}_0(v) - (1-3 \cos^2\alpha) {\cal J}_1(v)\Big] \Big({\rm C}_{\rm s}(g)+\frac{1}{2}\Big)\nonumber \\
&+&{\cal G}_0(v)-(1 - 3 \cos^2\alpha){\cal G}_1(v)\bigg\}\,,
\label{finsoft}
\ea
where ${\cal G}_1(v)$ is a complicated function expressed as a one-dimensional integral that requires numerical evaluation. Its explicit form is given by
\be
{\cal G}_1(v) =\frac{12v^{3}-7v}{35}+{\cal G}_1^{(a)}(v)+{\cal G}_1^{(b)}(v)+{\cal G}_1^{(c)}(v)+{\cal G}_1^{(d)}(v)\,,
\ee
with
\ba
{\cal G}_1^{(a)}(v)&=&\frac{1}{6\pi v^4}
\int_{0}^{v}\, d \wh\,\frac{\wh}{1-\hat{\omega}^{2}}\bigg\{2\Big[-9\wh^{4}+(13+3 v^{2})\hat{\omega}^{2}-7v^2\Big]\Theta_L(\hat{\omega})-\frac{v^2-\wh^2}{1-\hat{\omega}^{2}}\nonumber \\ 
&\times&\Big[21 \wh^4 - (8 + 9 v^2) \wh^2 + 6 v^2 - 4\Big]\Theta_T(\hat{\omega})\bigg\}\,,\\ 
{\cal G}_1^{(b)}(v)&=&\frac{1}{8v^4}
\int_{0}^{v}\,d\hat{\omega}\,\hat{\omega}^{2}
\bigg\{4\Big[17\hat{\omega}^{4}-(8+11v^{2})\hat{\omega}^{2}+4v^{2}\Big]{\rm R}_L(\hat{\omega})+\frac{v^2-\wh^2}{1-\hat{\omega}^{2}}\nonumber \\
&\times&\Big[37 \wh^4 - (35 + 9 v^2) \wh^2 + 7 v^2 + 4\Big]{\rm R}_T(\hat{\omega})\bigg\}\,,\\ 
{\cal G}_1^{(c)}(v)&=&\frac{16}{3\pi v^{4}}
\int_{0}^{v}\,d\wh\,\frac{\wh\,\big(v^{2}-\wh^{2}\big)}{{\cal N}(\wh)\,\big(1-\wh^{2}\big)}
\Big[\Theta_{L}(\wh)-2\,\Theta_{T}(\wh)\Big]\,,\\ 
{\cal G}_1^{(d)}(v)&=&\frac{8}{v^{4}}
\int_{0}^{v}\,d\wh\,\frac{\wh^2 \big(v^{2}-\wh^{2}\big)}{{\cal N}(\wh)}\Big[\big({\rm R}_L(\wh)-{\rm R}_{T}(\wh)\big)-2\big({\rm Q}_L(\wh)\Theta_L(\wh)-{\rm Q}_T(\wh)\Theta_T(\wh)\big)\Big]\,,\nonumber \\
\ea
and
\be
{\cal N}(\wh)=9\pi^{2}\wh^{2}\left(1-\wh^{2}\right)^{2}+\left(4-6\wh^{2}-3\wh\left(1-\wh^{2}\right)\ln\frac{1+\wh}{1-\wh}\right)^{2}\, .
\ee

For later use, we also provide the $v \rightarrow 0$ asymptotic behaviors of the above-defined functions involved in the energy loss, as follows:
\be
{\cal J}_0(v) \to \frac{2}{3}v\, ,\quad {\cal F}_0(v) \to \frac{2}{3}v \ln 2 \, ,\quad {\cal G}_0(v) \to \frac{1}{3}v\, ,
\ee
and
\be
{\cal J}_1(v) \to -\frac{2}{15}v \, ,\quad  {\cal F}_1(v) \to \frac{1}{5} v\left(1-\frac{2}{3}\ln 2\right)\, ,\quad  {\cal G}_1(v) \to -\frac{1}{3}v\, .
\ee

\section{results and discussions}
\la{numeric}

We now summarize the main results of the above calculations for the collisional energy loss in a local and quasi-static anisotropic background\footnote{For an anisotropic QGP that possesses unstable modes~\cite{Romatschke:2003ms,Romatschke:2004jh}, the complete energy exchange may also depend on the initial field and its time evolution~\cite{Mrowczynski:2016etf}. This lies beyond the scope of the present work.}. From Eqs.~(\ref{divhard}) and (\ref{divsoft}), it is evident that in the weak-coupling limit $g T\ll q^* \ll T$, the cancellation of the artificial cutoff $q^*$ between the hard and soft contributions also holds in the presence of a small anisotropy. Furthermore, the finite part given by Eqs.~(\ref{finhard}) and (\ref{finsoft}) represents the anisotropy correction to the isotropic/equilibrium collisional energy loss. By combining this result with Eqs.~(\ref{finhard0}) and (\ref{qgfinsoft0}), the total energy loss to linear order in $\xi$ is obtained as
\be\label{final}
-{\Big(\frac{dE}{dx}\Big)}=-{\Big(\frac{dE}{dx}\Big)}_{\rm {(0)}}-{\Big(\frac{dE}{dx}\Big)}_{\rm {(1)}}\,,
\ee
with
\be\label{reun}
-{\Big(\frac{dE}{dx}\Big)}_{\rm {(0)}}=\frac{g^2m_D^2}{6\pi} {\cal A}_0(v,g)\,,\quad  -{\Big(\frac{dE}{dx}\Big)}_{\rm {(1)}}=-\frac{g^2m_D^2}{6\pi}\frac{\xi}{3}\left[ {\cal A}_0(v,g)-(1-3\cos^2\alpha){\cal A}_1(v,g)\right]\,.
\ee
and
\be
 {\cal A}_i(v,g) ={\cal J}_i(v)\big({\rm C}_{\rm h}-{\rm C}_{\rm s}(g)\big) + {\cal F}_i(v)- {\cal G}_i(v)\,,\quad {\rm for}\quad i=0,1\,.
\ee

In the above result, the anisotropy correction $-\big(d E/d x\big)_{\rm (1)}$ can be decomposed into a monopole (angle-independent) part and a quadrupole part proportional to the second Legendre polynomial $P_2(\cos\alpha)=-(1-3\cos^2\alpha)/2$. The monopole contribution is simply proportional to $-\big(d E/d x\big)_{\rm (0)}$ and suppresses the isotropic energy loss. The quadrupole contribution, which contains the angular dependence of the anisotropy correction, vanishes after averaging over all possible angles between ${\bf v}$ and ${\bf n}$. It also vanishes for the special incident angle satisfying $\cos^2 \alpha=1/3$, i.e., $\alpha\approx 55^\circ$. In both cases, the collisional energy loss is simply given by isotropic energy loss multiplied by a factor of $1-\xi/3$, indicating a reduced energy loss in an anisotropic QGP. In addition, owing to the $\cos^2\alpha$ angular dependence, the anisotropic energy loss is symmetric under $\alpha \to -\alpha$ and $\alpha \to \pi-\alpha$. Therefore, it suffices to consider $0 \le \alpha \le \pi/2$. Notice that $-\big(d E/d x\big)$ is a monotonic function of $\alpha$ in the range $0 \le \alpha \le \pi/2$, implying that the energy loss reaches its extreme values when the incident velocity is parallel or perpendicular to the direction of anisotropy. 

Furthermore, when varying the incident angle of the heavy quark, the sign of ${\cal A}_1(v,g)$ determines the minimum and maximum of the anisotropic energy loss. For ${\cal A}_1(v,g)>0$, the minimum of $-\big(d E/d x\big)$ occurs when ${\bf v} \parallel {\bf n}$, whereas ${\bf v} \perp {\bf n}$ corresponds to the maximum. For ${\cal A}_1(v,g)<0$, the situation is reversed. However, ${\cal A}_1(v,g)$ can take either sign, depending not only on the incident velocity $v$ but also on the coupling $g$. In the weak-coupling limit $g\ll 1$, where our calculation becomes theoretically more reliable, we have
\be\label{a1lim}
{\cal A}_1(v,g) \xrightarrow{g\ll 1} - {\cal J}_1(v) \ln g \,.
\ee
Since ${\cal J}_1(v)<0$ for $0<v<1$, we find that a heavy quark loses more energy when propagating parallel to the direction of anisotropy, compared to the case where ${\bf v} \perp {\bf n}$. Notice that in the weak-coupling limit, the angular dependence is most pronounced at $v\approx 0.79$, where $|{\cal J}_1(v)|$ reaches its maximum.

At relatively large couplings, the sign of ${\cal A}_1(v,g)$ depends nontrivially on both $v$ and $g$, so that there is no simple answer to the question of which incident angle maximizes or minimizes the anisotropic energy loss. As we have just shown, ${\cal A}_1(v,g)$ is significantly negative for very small $g$. On the other hand, since $\partial {\cal A}_1(v,g)/\partial g=-{\cal J}_1(v)/g>0$, ${\cal A}_1(v,g)$ increases monotonically with $g$ and eventually crosses zero, becoming positive. For a given $v$, we can define $g_{0}(v)$ as the coupling at which the angular dependence of the energy loss vanishes, i.e., ${\cal A}_1(v,g)=0$. It is found that
\be\label{g0v}
g_{0}(v)=\frac{1}{\sqrt{1+N_f/6}} e^{({\cal F}_1(v)- {\cal G}_1(v))/{\cal J}_1(v)+{\rm C_h}}\, .
\ee
It can be shown that $g_{0}(v)$ increases rapidly with the
incident velocity and becomes very large as $v\rightarrow 1$. However, the energy loss formula in Eq.~(\ref{final}) ceases to be physically meaningful at large couplings, where it would yield a negative isotropic energy loss\footnote{This is just because the approximation $g T\ll q^* \ll T$ is no longer valid, but does not mean a physical energy gain.}. Therefore, ${\cal A}_0(v,g)$ must be strictly positive, which imposes the following constraint on the coupling constant
\be
g< \frac{1}{\sqrt{1+N_f/6}} e^{({\cal F}_0(v)- {\cal G}_0(v))/{\cal J}_0(v)+{\rm C_h}}\, .
\ee
Because the right-hand side of this inequality is an increasing function of $v$, to ensure a positive $-\big(dE/dx\big)_{\rm (0)}$
for arbitrary incident velocity, the constraint must be imposed in the limit $v\rightarrow 0$, where the right-hand side attains its minimum.
Using $\lim_{v\rightarrow 0}\big[ {\cal F}_0(v)- {\cal G}_0(v)\big]/{\cal J}_0(v)=\ln 2-1/2$, one can show that the value of $g$ cannot exceed $\sim 1.08$ for three flavors (we adopted $N_f=3$ in the numerical evaluations). In what follows, we restrict ourselves to this moderately small coupling regime $g< 1$. It is also worth noting that, strictly speaking, the energy loss given by Eq.~(\ref{final}) is applicable only in the weak-coupling limit. For relatively large couplings near unity, the results should be interpreted as an extrapolation of the weak-coupling expression.

\begin{figure*}[t]
    \begin{center}
    \includegraphics[width=0.6\linewidth]{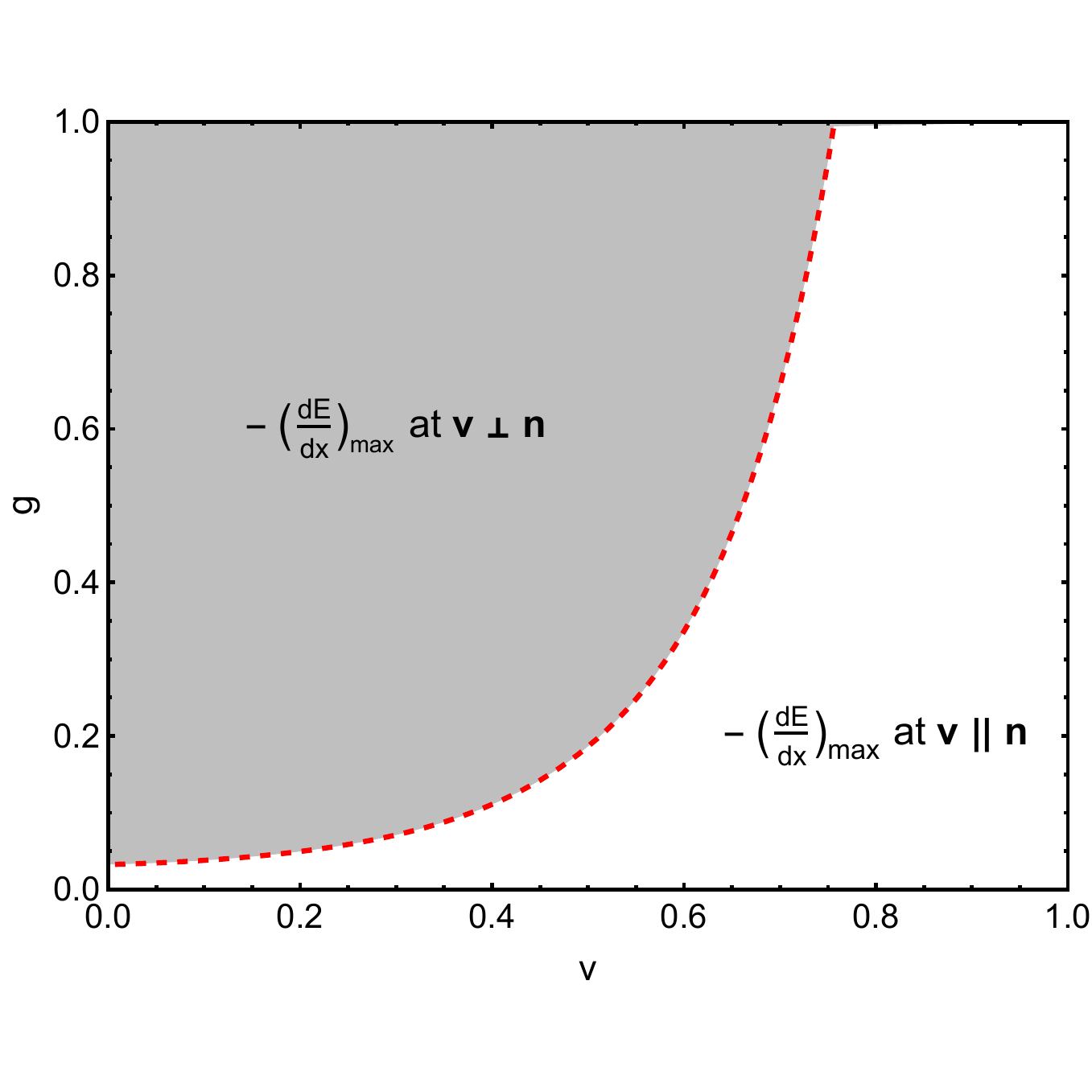}
    \end{center}
    \vspace{-5mm}
    \caption{The boundary curve in the $v\text{--}g$ plane. The upper-left region corresponds to the maximum of $-\big(dE/dx\big)$ appearing at ${\bf v} \perp {\bf n}$ and the lower-right region to that appearing at ${\bf v} \parallel {\bf n}$.}
    \label{reg1}
\end{figure*}

Numerically, we find that $g_0(v)$ is larger than $1$ when $v> 0.76$. Consequently, for any physically meaningful coupling, a heavy quark with $v>0.76$ loses more energy when ${\bf v}\parallel{\bf n}$. The same conclusion also holds, regardless of the magnitude of the
incident velocity, for couplings below $\sim 0.03$. This is because the minimum of $g_0(v)$ occurs as $v\to 0$, where one can analytically show that $g_0(v\to 0)=2 e^{{\rm C_h}-4}/\sqrt{1+N_f/6}\approx 0.03$. On the other hand, if $v$ and $g$ satisfy both $v < 0.76$ and $0.03 < g < 1$, the conclusion depends on their particular values. For $g<g_{0}(v)$, ${\cal A}_1(v,g)$ is negative and the maximum of $-\big(d E/d x\big)$ appears at ${\bf v} \parallel {\bf n}$; while for $g>g_{0}(v)$, the positive ${\cal A}_1(v,g)$ indicates that $-\big(d E/d x\big)$ attains the maximum at ${\bf v} \perp {\bf n}$. In Fig.~\ref{reg1}, the upper-left region in the $v\text{--}g$ plane indicates that the energy loss is maximal in the case of perpendicular propagation, whereas the lower-right region corresponds to the case where $-\big(dE/dx\big)$ becomes largest when ${\bf v}$ is parallel to ${\bf n}$. The boundary curve separating the two regions is determined by $g_0(v)$ and terminates at $(0.76,1)$ and $(0,0.03)$.

Once anisotropy correction is taken into account, it is also very interesting to ask whether the energy loss increases or decreases relative to the isotropic case. In the weak-coupling limit, a similar result as given in Eq.~(\ref{a1lim}) also holds for ${\cal A}_0(v,g)$,
\be\label{a0lim}
{\cal A}_0(v,g) \xrightarrow{g\ll 1} - {\cal J}_0(v) \ln g \,.
\ee
Therefore, the anisotropy correction can be obtained as
\ba
-{\Big(\frac{dE}{dx}\Big)}_{\rm {(1)}} &\xrightarrow{g\ll 1}&  \xi \frac{g^2 m_D^2}{18\pi} \left[{\cal J}_0(v)-(1-3\cos^2\alpha){\cal J}_1(v)\right] \ln g\nonumber \\
&\le& \xi \frac{g^2 m_D^2}{18\pi} \left[{\cal J}_0(v)+2 {\cal J}_1(v)\right] \ln g\,.
\ea
We find that since ${\cal J}_0(v)+2 {\cal J}_1(v)>0$, regardless of the incident angle, the energy loss in an anisotropic QGP is always reduced compared to the isotropic case.

The situation becomes more involved for moderately small couplings, since the sign of the anisotropy correction is now determined jointly by $v$, $g$, and $\alpha$. We point out that, independently of the incident angle $\alpha$, the quantity $-\left[ {\cal A}_0(v,g)-(1-3\cos^2\alpha){\cal A}_1(v,g)\right]$ increases monotonically with the coupling, because ${\cal J}_0(v)-(1-3\cos^2\alpha){\cal J}_1(v)>0$. As a result, for fixed $v$ and $\alpha$, this quantity is maximal at the largest coupling, $g=1$. It can be shown that to get a negative anisotropy correction for all $0<g<1$, the incident angle and velocity should satisfy the following conditions\footnote{${\cal A}_0(v,1)$ is always positive. For ${\cal A}_1(v,1)$, $v\approx0.76$ is a turning point: ${\cal A}_1(v,1)>0$ for $0<v<0.76$, while ${\cal A}_1(v,1)<0$ for $v>0.76$.}
\be\label{con1}
3\cos^2\alpha-1>-{\cal A}_0(v,1)/{\cal A}_1(v,1) \quad {\rm for}\quad  0<v<0.76\, ,
\ee
and
\be\label{cond2}
3\cos^2\alpha-1<-{\cal A}_0(v,1)/{\cal A}_1(v,1) \quad {\rm for} \quad 0.76<v<1\, .
\ee
For $0.76<v<1$, we numerically find that the minimum of $-{\cal A}_0(v,1)/{\cal A}_1(v,1)$ is about $7.79$ which is obtained at $v\approx 0.91$. Therefore, Eq.~(\ref{cond2}) can be satisfied for arbitrary $\alpha$.  For $0<v<0.76$, the maximum of $-{\cal A}_0(v,1)/{\cal A}_1(v,1)$ appears as $v\rightarrow 0$ and is given by $5+35/\big[2({\rm C}_{\rm h}-{\rm C}_{\rm s}(1))+2\ln 2-8\big]\approx -0.11$.  Consequently, there exists a critical incident angle $\alpha_{\rm c}\approx 57^\circ$ below which\footnote{Due to the symmetry, the same is true for $123^\circ <\alpha<\pi$.} the energy loss in an anisotropic plasma is reduced relative to the isotropic energy loss. This conclusion holds for any incident velocity and physically meaningful coupling $g< 1$. Furthermore, we find that for $0<v<0.76$, $-{\cal A}_0(v,1)/{\cal A}_1(v,1)$ decreases with increasing $v$. By setting $-{\cal A}_0(v,1)/{\cal A}_1(v,1)=-1$ in Eq.~(\ref{con1}), we can identify a critical incident velocity $v_{\rm c}\approx 0.50$, above which the anisotropic case exhibits a reduced energy loss for any $\alpha$ and $g<1$.



In fact, there also exits a critical coupling constant $g_{\rm c}\approx 0.60$. For couplings smaller than this critical value, we find a reduced energy loss in an anisotropic plasma which holds regardless of the incident angle and velocity. To get a negative correction to the isotropic energy loss, the coupling $g$ should satisfy the following condition
\be
\ln g<{\rm C}_{\rm h}-\ln\sqrt{1+N_f/6}+\frac{ {\cal F}_0(v)- {\cal G}_0(v)-(1-3\cos^2\alpha)({\cal F}_1(v)- {\cal G}_1(v))}{{\cal J}_0(v)-(1-3\cos^2\alpha){\cal J}_1(v)}\,.
\ee
Further investigation shows that the right-hand side of this inequality attains its minimum when $\alpha=\pi/2$ and $v\rightarrow 0$, therefore, we can obtain the critical coupling constant as $g_{\rm c}=2 e^{{\rm C_h}-13/12}/\sqrt{1+N_f/6}\approx 0.60$.

\begin{figure*}[t]
    \begin{center}
    \includegraphics[width=0.6\linewidth]{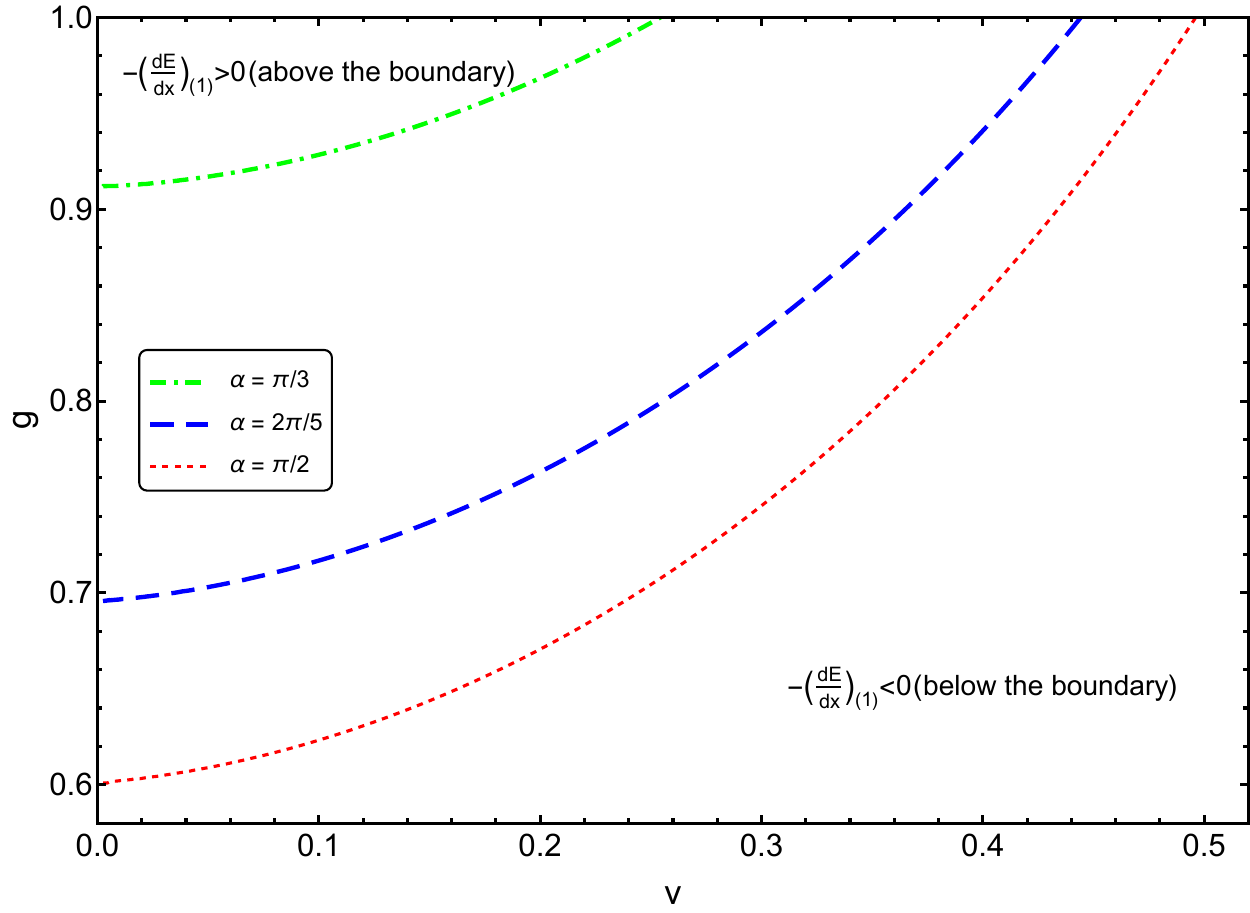}
    \end{center}
    \vspace{-5mm}
    \caption{The boundary curves in the $v\text{--}g$ plane for several fixed incident angles. For each curve, the upper/lower region corresponds to enhancement/suppression of the isotropic energy loss due to the unnormalized anisotropy correction.}
\label{av}
\end{figure*}

More generally, at several fixed incident angles, Fig.~\ref{av} shows the boundary curves along which the anisotropy correction vanishes. These curves divide the $v\text{--}g$ plane into two regions: the upper region corresponds to enhancement of the isotropic energy loss due to the anisotropy correction, while the lower-left region corresponds to suppression. Furthermore, the boundary curves terminate on both axes. Their vertical-axis endpoints are determined by $g= 2 e^{{\rm C_h}-4+35/(12-6\cos^2\alpha)}/\sqrt{1+N_f/6}$. Explicitly, as the incident angle decreases from large to small, the corresponding values of $g$ are $0.60$, $0.70$, and $0.91$. Accordingly, the horizontal-axis endpoints are numerically found to be located at $v=0.50$, $0.44$ and $0.25$. The results presented in this figure also indicate that for any pair $(v,g)$ lying below the short-dashed boundary curve (for $\alpha=\pi/2$), the anisotropic correction is negative for all incident angles. 

\begin{figure*}[t]
    \begin{center}
    \includegraphics[width=0.6\linewidth]{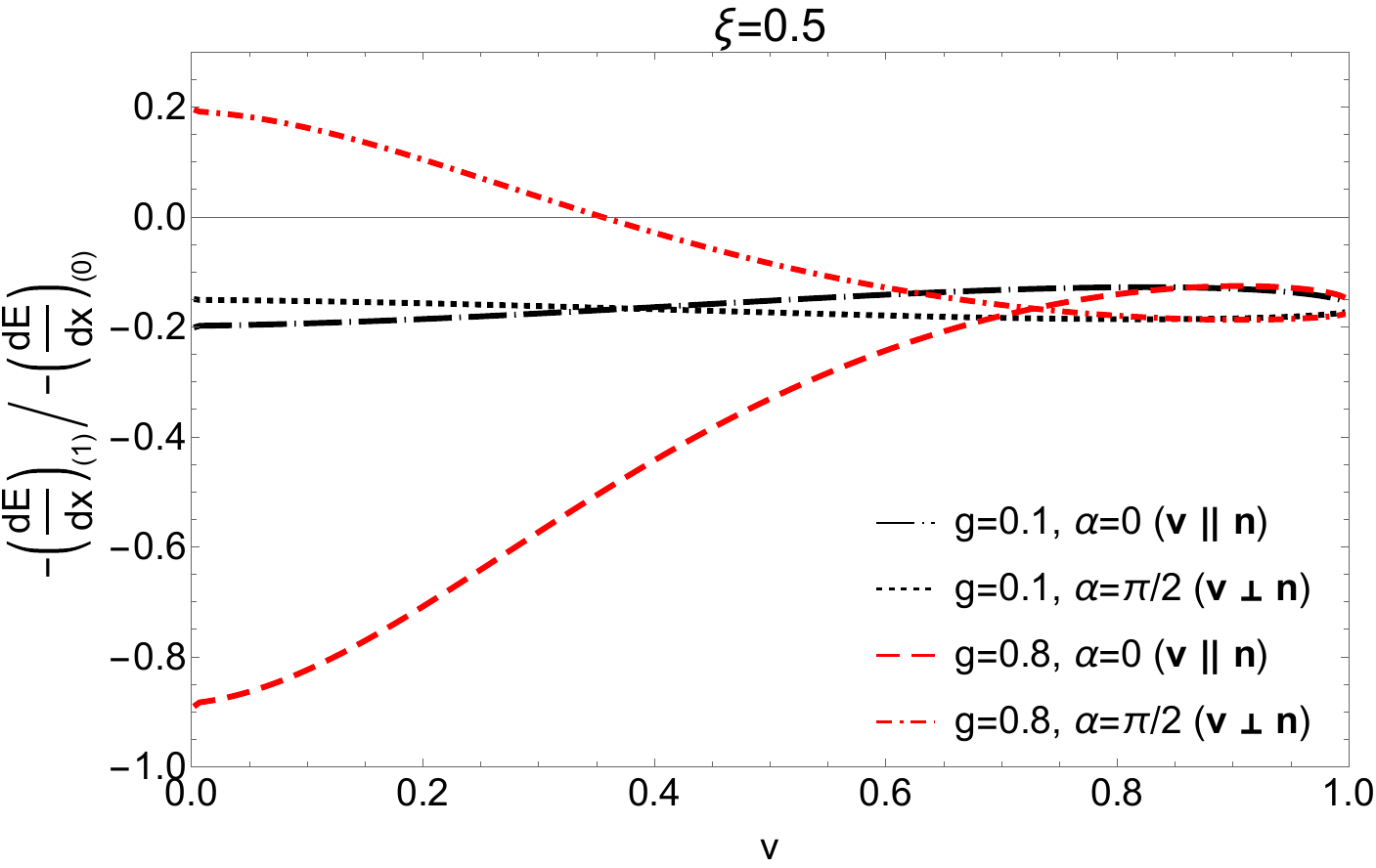}
    \end{center}
    \vspace{-5mm}
    \caption{The ratio of the anisotropy correction to the isotropic energy loss as a function of $v$ for different $g$ and $\alpha$. The results are obtained based on the unnormalized distribution function and the anisotropy parameter $\xi=0.5$.}
\label{ela}
\end{figure*}

To better illustrate the effect of anisotropy on the energy loss, we show in Fig.~\ref{ela} the relative anisotropy correction, which is defined as the ratio of the anisotropy correction to the isotropic energy loss, as a function of the incident velocity. We choose $\xi=0.5$ and consider two special incident directions ${\bf v}\parallel{\bf n}$ and ${\bf v}\perp{\bf n}$. Indeed, in the weak-coupling limit, this ratio is given by 
\be
\frac{-\big(dE/dx\big)_{(1)}}{-\big(dE/dx\big)_{(0)}}\xrightarrow{g\ll 1}-\frac{\xi}{3} \left[1-(1-3\cos^2\alpha)\frac{{\cal J}_1(v)}{{\cal J}_0(v)}\right]\, .
\ee
As both $\alpha$ and $v$ vary, the range of the above ratio is $(-2\xi/5,-\xi/5)$\footnote{This follows from the fact that ${\cal J}_1(v)/{\cal J}_0(v)$ is negative and attains a minimum of $-1/5$ in the limit $v \to 0$.}. On the other hand, for small coupling $g=0.1$ as shown in this figure, the anisotropy correction suppresses the energy loss, which depends only moderately on $v$. We also find that $-\big(d E/d x\big)$ at ${\bf v}\parallel{\bf n}$ is larger than that at ${\bf v}\perp{\bf n}$ when $v>0.38$, which is consistent with the result in Fig.~\ref{reg1}. For a relatively large coupling $g=0.8$, the anisotropy correction becomes significant at small velocities where the energy loss is enhanced for perpendicular incidence and reduced for parallel incidence. Furthermore, in accordance with our previous discussions, Fig.~\ref{ela} also demonstrates that the anisotropy correction flips sign at $v\approx 0.35$ for ${\bf v}\perp{\bf n}$, and the energy loss with ${\bf v}\parallel{\bf n}$ starts to exceed that with ${\bf v}\perp{\bf n}$ when $v>0.73$. Finally, we point out that as $v\to 1$, the isotropic energy loss diverges, while ${\cal A}_1(1,g)$ remains finite. Therefore, this ratio approaches $-\xi/3$ for any coupling and incident angle.

\section{The anisotropy correction with fixed number density or energy density}
\la{fixed}

In general, the comparison of physical quantities between isotropic and anisotropic QGP can be also carried out by fixing the number density or the energy density of the two systems. In this section, we examine the heavy-quark collisional energy loss in anisotropic QGP and the corresponding modification with respect to the isotropic case, under the condition that either the number density or the energy density is fixed.

\subsection{The anisotropy correction with fixed number density}
\la{fix n}

For the anisotropic distribution function given in Eq.~(\ref{anisodis}), the reduction of the collisional energy loss in an anisotropic QGP can be partly attributed to the reduced number density. To keep the number density the same in both the isotropic and anisotropic cases, one may also include a normalization factor $\sqrt{1+\xi}$ in Eq.~(\ref{anisodis}). As a result, the anisotropy correction due to hard scatterings receives an extra contribution which is simply given by
\be
-{\Big(\frac{dE}{dx}\Big)}_{\rm {hard(1)}}^{{\rm ex.}}= -\frac{\xi}{2} {\Big(\frac{dE}{dx}\Big)}_{\rm {hard(0)}}\, ,
\label{exhard}
\ee
where ${\big(dE/dx\big)}_{\rm {hard(0)}}$ is given by the sum of Eqs.~(\ref{divhard0}) and (\ref{finhard0}). Notice that this extra contribution has no dependence on the angle $\alpha$. For the soft scatterings, the inclusion of the normalization factor modifies the structure functions. To linear order in $\xi$, the modification results in 
\be
\alpha_{(1)}\rightarrow \alpha_{(1)}+\frac{1}{2} \Pi_T(\wh)\,\qquad \beta_{(1)}\rightarrow \beta_{(1)}+\frac{1}{2} \wh^2 \Pi_L(\wh)\,,
\ee
while $\gamma_{(1)}$ and $\delta_{(1)}$ remain unchanged. Accordingly, the following extra contribution should be added to $\Delta_{(1)}^{ij}(Q)$ given in Eq.~(\ref{D1}),
\be
\left(\Delta^{\rm ex.}\right)_{(1)}^{ij}(Q)=-\frac{m_D^2}{2}\bigg[\frac{\Pi_T(\wh)}{\big(q^{2}-\omega^{2}+m_D^2\,\Pi_T(\wh)\big)^2}A^{ij}+\frac{\Pi_L(\wh)}{\wh^2\big(m_D^2 \Pi_L(\wh)-q^2\big)^2}B^{ij}\bigg]\,. 
\ee
It can be shown that the following relation holds to linear order in $\xi$,
\be
v^i\big[{\tilde \Delta}^{ij}_{(0)}(\xi,Q)
-{(\Delta_{\rm bare})}^{ij}(Q)\big]v^j=v^i\big[ \Delta^{ij}_{(0)}(Q)
-{(\Delta_{\rm bare})}^{ij}(Q)\big]v^j+\xi v^i \left(\Delta^{\rm ex.}\right)_{(1)}^{ij}(Q)v^j\, .
\ee
Here, the tilde-marked isotropic propagator  ${\tilde \Delta}^{ij}_{(0)}(\xi,Q)$ is defined from its untilded counterpart given in Eq.~(\ref{diso}), with the gluon self-energies  $\Pi_T(\wh)$ and $\Pi_L(\wh)$ replaced by
\be
{\tilde \Pi}_T(\wh)=\sqrt{1+\xi}\,\Pi_T(\wh)\,,\qquad {\tilde \Pi}_L(\wh)=\sqrt{1+\xi}\,\Pi_L(\wh)\,,
\ee
respectively. Based on this relation, we can further show that
\ba\label{ex2}
-{\Big(\frac{dE}{dx}\Big)}_{\rm {soft(0)}}-{\Big(\frac{dE}{dx}\Big)}_{\rm {soft(1)}}^{{\rm ex.}}&=&- \frac{g^2 m_D^2}{6\pi^2 v^2}{\rm Im} \int_{-v}^v d \wh\, \wh \Bigg\{{\tilde \Pi}_L(\wh) \ln\left(1-\frac{(q^*/m_D)^2}{{\tilde \Pi}_L(\wh)}\right) \nonumber \\
&+&\frac{v^2-\hat{\omega}^2}{(1-\hat{\omega}^2)^2}
{\tilde\Pi}_T(\wh) \ln\left(1+\frac{(q^*/m_D)^2(1-\hat{\omega}^2)}{{\tilde \Pi}_T(\wh)}\right)\Bigg\} \, . 
\ea
The right-hand side of the above equation is formally identical to the isotropic soft contribution to the energy loss, except that $\Pi_{T/L}(\wh)$ is replaced by ${\tilde \Pi}_{T/L}(\wh)$. This equation is also valid to linear order in $\xi$. After taking into account the weak-coupling limit $g T\ll q^*$, we obtain the following explicit result for the extra terms in the anisotropy correction
\be
-{\Big(\frac{dE}{dx}\Big)}_{\rm {soft(1)}}^{{\rm ex.}}= -\frac{\xi}{2} {\Big(\frac{dE}{dx}\Big)}_{\rm {soft(0)}}- \xi \frac{g^2  m_D^2 }{24 \pi} {\cal J}_0(v)\, , 
\label{exsoft}
\ee
where the first and second terms on the right-hand side come from the anisotropy correction to the gluon self-energy outside and inside the logarithm in Eq.~(\ref{ex2}), respectively.
Just like the hard scatterings, the above extra contribution also has no dependence on the angle $\alpha$. Furthermore, combining Eq.~(\ref{exhard}) and Eq.~(\ref{exsoft}), it is clear that the cutoff dependence in the extra terms cancels between the hard and soft contributions. This indicates that the collisional energy loss in a QGP with small anisotropy is independent of $q^*$, regardless of whether or not the number density is fixed.

To linear order in $\xi$, the total collisional energy loss at fixed number density can be written as
\ba\label{reno}
-{\Big(\frac{dE}{dx}\Big)}^{\mathrm{norm}}&=&\frac{g^2m_D^2}{6\pi}\bigg\{\Big(1+\frac{\xi}{2}\Big){\cal A}_0(v,g)-\frac{\xi}{3}\left[{\cal A}_0(v,g)-(1-3\cos^2\alpha){\cal A}_1(v,g)\right]-\frac{\xi}{4}{\cal {J}}_0(v)\bigg\}\nonumber \\
&=&\frac{g^2m_D^2}{6\pi}\left\{{\cal A}_0(v,g)-\frac{\xi}{3}\left[{\cal A}_0^{\prime}(v,g)-(1-3\cos^2\alpha){\cal A}_1(v,g)\right]\right\}\,,
\ea
with
\be
{\cal A}_0^{\prime}(v,g)\equiv-\frac{1}{2}{\cal A}_0(v,g)+\frac{3}{4}{\cal J}_0(v)\, .
\ee
As we can see, the quadrupole angular structure is identical to that of the unnormalized result in Eq.~(\ref{final}). Hence, the earlier results for the unnormalized case as shown Fig.~\ref{reg1} carry over to the fixed-number-density case. 

On the other hand, the sign of the anisotropy correction shows a different behavior compared to the unnormalized case. Such a correction reads
\be\label{nd3}
-{\Big(\frac{dE}{dx}\Big)}_{\rm {(1)}}^{\rm{norm}}=-\frac{g^2m_D^2}{6\pi}\frac{\xi}{3}\left[{\cal A}_0^{\prime}(v,g)-(1-3\cos^2\alpha){\cal A}_1(v,g)\right]\, .
\ee
Although ${\cal A}_0(v,g)$ is strictly positive for $0<g<1$, the sign of ${\cal A}_0^{\prime}(v,g)$ depends on both $v$ and $g$. Therefore, when the quadrupole contribution vanishes, one cannot simply expect a reduction in the energy loss from the anisotropy correction. 

To check the sign of Eq.~(\ref{nd3}), we first consider the weak-coupling limit where 
\be\label{a0plim}
{\cal A}^\prime_0(v,g) \xrightarrow{g\ll 1}  \frac{1}{2} {\cal J}_0(v) \ln g \,.
\ee
Consequently, the anisotropy correction can be written as
\ba
-{\Big(\frac{dE}{dx}\Big)}_{\rm {(1)}}^{\rm{norm}} &\xrightarrow{g\ll 1} &- \xi \frac{g^2 m_D^2}{36\pi} \left[{\cal J}_0(v)+2(1-3\cos^2\alpha){\cal J}_1(v)\right] \ln g \nonumber \\
&\ge& - \xi \frac{g^2 m_D^2}{36\pi} \left[{\cal J}_0(v)+2 {\cal J}_1(v)\right] \ln g >0\,,
\ea
which indicates that this correction enhances the energy loss, in contrast to the unnormalized case.

For moderately small couplings, the quantity $-\left[{\cal A}_0^{\prime}(v,g)-(1-3\cos^2\alpha){\cal A}_1(v,g)\right]$ decreases monotonically as the coupling increases and attains its minimum at $g=1$. Accordingly, to ensure that the anisotropy correction is positive for all $0<g<1$, the following conditions must be satisfied,
\be\label{con3}
3\cos^2\alpha-1<-{\cal A}^\prime_0(v,1)/{\cal A}_1(v,1) \quad {\rm for}\quad  0<v<0.76\, ,
\ee
and
\be\label{cond4}
3\cos^2\alpha-1>-{\cal A}^\prime_0(v,1)/{\cal A}_1(v,1) \quad {\rm for} \quad 0.76<v<1\, .
\ee
For $0<v<0.76$, the minimum of $-{\cal A}^\prime_0(v,1)/{\cal A}_1(v,1)$ occurs as $v\rightarrow 0$ and is found to be $[10-5\ln 2-5 ({\rm C}_{\rm h}-{\rm C}_{\rm s}(1))]/[2\ln 2-8+2 ({\rm C}_{\rm h}-{\rm C}_{\rm s}(1))]\approx -1.04$. Consequently, the first inequality cannot be satisfied, indicating that no critical incident angle exists as defined in the unnormalized case. On the other hand, analyzing the $v$-dependent behavior of $-{\cal A}^\prime_0(v,1)/{\cal A}_1(v,1)$ in Eq.~(\ref{cond4}) allows us to identify a critical incident velocity $v_{\rm c}\approx 0.98$; above this value, the anisotropic case exhibits an enhanced energy loss for all $\alpha$ and $g<1$. Similarly, to get a positive correction to the isotropic energy loss, the coupling $g$ should satisfy the following condition
\be
\ln g<{\rm C}_{\rm h}-\ln\sqrt{1+N_f/6}+\frac{ 2[{\cal F}_0(v)- {\cal G}_0(v)]+4(1-3\cos^2\alpha)({\cal F}_1(v)- {\cal G}_1(v))-3 {\cal J}_0(v)}{2{\cal J}_0(v)+4(1-3\cos^2\alpha){\cal J}_1(v)}\,.
\ee
Over the region $0<v<1$ and $0\le \alpha\le\pi/2$, the minimum of the right-hand side of this inequality is $2 e^{{\rm C_h}-26/9}/\sqrt{1+N_f/6}$, attained in the limit $v\rightarrow 0$ and $\alpha=0$. The critical coupling constant is therefore given by $g_{\rm c}\approx 0.10$; below this value, the anisotropic case exhibits an enhanced energy loss for all $\alpha$ and $v$. 

\begin{figure*}[t]
    \begin{center}
    \includegraphics[width=0.6\linewidth]{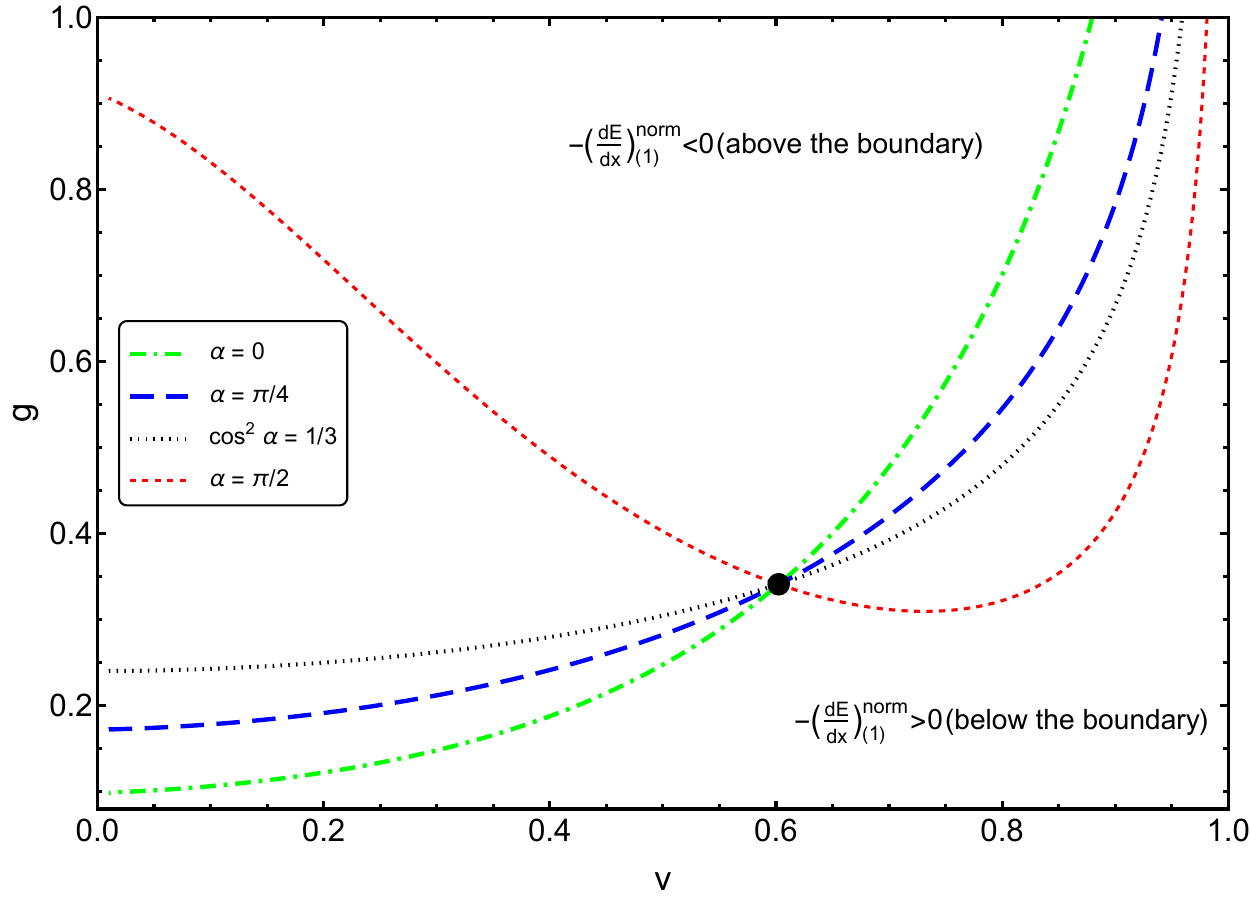}
    \end{center}
    \vspace{-5mm}
    \caption{The boundary curves in the $v\text{--}g$ plane for several fixed incident angles. For each curve, the upper/lower region corresponds to suppression/enhancement of the isotropic energy loss due to the anisotropy correction with fixed-number-density.}
\label{av2}
\end{figure*}

Fig.~\ref{av2} shows the corresponding boundary curves in the fixed-number-density case. The upper and lower regions correspond to suppression and enhancement of the isotropic energy loss, respectively. This is exactly opposite to the unnormalized case. Furthermore, the endpoints along the vertical-axis are determined by $g= 2 e^{{\rm C_h}-4+10/(3+6\cos^2\alpha)}/\sqrt{1+N_f/6}$, corresponding to $0.91$, $0.24$, $0.17$, and $0.10$ for different incident angles presented in this figure. The horizontal-axis endpoints are numerically found to be located at $v=0.98$, $0.96$, $0.94$, and $0.88$. Notice that for $\cos^2\alpha=1/3$, the quadrupole part in the anisotropy correction vanishes. However, unlike the unnormalized case where the anisotropy suppresses the energy loss, $-\big(d E/d x\big)^{\rm {norm}}_{{\rm (1)}}$ can be either positive or negative, depending on both $v$ and $g$.

Another notable difference between Figs.~\ref{av} and \ref{av2} is that the boundary curves in Fig.~\ref{av} rise monotonically with increasing $v$, whereas in Fig.~\ref{av2}, for large incident angles (specifically, when $\alpha$ exceeds about $\pi/3$), the boundary curves exhibit non-monotonic behavior as a function of $v$. In the former case, each curve attains its minimum $g_{\rm min}$ at $v=0$, i.e., $g_{\rm min}=g(v=0)$. For any given $g>g_{\rm min}$, the sign of the anisotropy correction reverses at a particular value of $v$. In the latter case, by contrast, the minimum occurs at $v\neq0$. For a coupling constant satisfying $g_{\rm min}<g<g(v=0)$, the sign of the anisotropy correction may reverse twice as the velocity increases from zero. We also find that in Fig.~\ref{av2}, all curves intersect at a special point, determined by ${\cal A}_1(v,g)={\cal A}_0^{\prime}(v,g)=0$. At this point, the anisotropy correction vanishes regardless of the incident angle. Numerically, this special point is given by $(0.60,0.34)$. In this figure, one can identify two nonsmooth boundary curves. One consists of the boundary curve for $\alpha=\pi/2$ with $v<0.60$ and the boundary curve for $\alpha=0$ with $v>0.60$; the other consists of the boundary curve for $\alpha=0$ with $v<0.60$ and the boundary curve for $\alpha=\pi/2$ with $v>0.60$. For a given pair $(v,g)$, the anisotropy correction is negative if $(v,g)$ lies above the first nonsmooth boundary curve and positive if it lies below the second one. In particular, in both cases the sign of the anisotropy correction is independent of the incident angle.

\begin{figure*}[t]
    \begin{center}
    \includegraphics[width=0.6\linewidth]{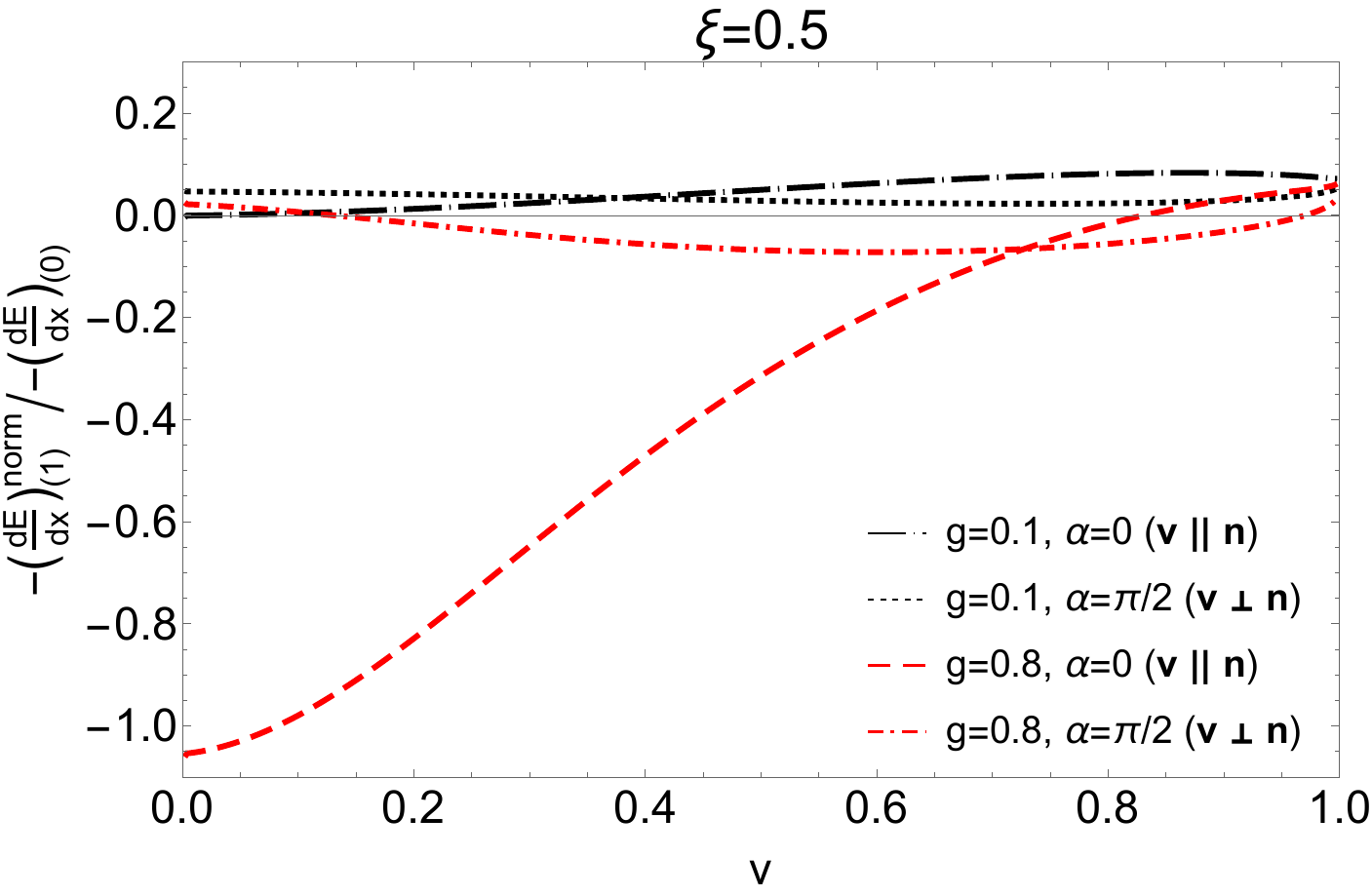}
    \end{center}
    \vspace{-5mm}
    \caption{The ratio of the anisotropy correction to the isotropic energy loss as a function of $v$ for different $g$ and $\alpha$. The results are obtained while keeping the number density equal to that of the isotropic medium and the anisotropy parameter $\xi=0.5$.}
\label{ra2}
\end{figure*}

For the fixed-number-density case, we also show the ratio of the anisotropy correction to the isotropic energy loss in Fig.~\ref{ra2}. It can be checked that the $v$-dependence of the ratio is actually consistent with the results presented in Figs.~\ref{reg1} and \ref{av2}. It should be particularly noted that, when $g=0.8$ and ${\bf v}\perp {\bf n}$, the ratio shown in the figure changes sign twice as $v$ varies. We also point out that when the heavy-quark propagation is parallel to the direction of anisotropy, the anisotropy correction is no longer quantitatively perturbative for relatively large coupling constants, e.g., $g=0.8$ as shown in this figure. In this case, one needs to consider even smaller anisotropy parameters to obtain reliable quantitative predictions. On the other hand, in the weak-coupling limit,
\be
\frac{-\big(dE/dx\big)^{\rm norm}_{(1)}}{-\big(dE/dx\big)_{(0)}}\xrightarrow{g\ll 1}\frac{\xi}{6} \left[1+2(1-3\cos^2\alpha)\frac{{\cal J}_1(v)}{{\cal J}_0(v)}\right]\, ,
\ee
which differs from the unnormalized case only by a shift of $\xi/2$, therefore, the range of the above ratio is $(\xi/10, 3\xi/10)$. Finally, we can also obtain the asymptotic value of this ratio as $v\rightarrow 1$, which is $\xi/6$ for any coupling and incident angle.

\subsection{The anisotropy correction with fixed energy density}
\la{fix e}

It is also useful to consider an alternative comparison in which the energy density of the anisotropic medium is constrained to be identical to that of the isotropic medium. As already discussed in Ref.~\cite{Zhao:2023mrz}, to fix the energy density, we can adjust the temperature-like scale $\lambda$ in the anisotropic distribution Eq.~(\ref{anisodis}) as the following
\be
\lambda\rightarrow T\left(\frac{1}{2(1+\xi)}+ \frac{\arctan{\sqrt{\xi}}}{2\sqrt{\xi}}\right)^{-\frac{1}{4}}=T\left(1+\frac{\xi}{6}\right)+{\cal O}(\xi^2)\, .
\ee
Interestingly, fixing the energy density with this adjustment also leaves the number density unchanged up to the linear order in $\xi$. Furthermore, such an adjustment only leads to a simple modification on the Debye mass, namely,
\be
m_D^2\rightarrow m_D^2\left(\frac{1}{2(1+\xi)}+ \frac{\arctan{\sqrt{\xi}}}{2\sqrt{\xi}}\right)^{-\frac{1}{2}}=m_D^2\left(1+\frac{\xi}{3}\right)+{\cal O}(\xi^2)\, .
\ee
As a result, the corresponding energy loss can be readily obtained from the unnormalized result by modifying the Debye mass according to the above equation. Clearly, such a modification does not affect the cancellation of the cutoff $q^*$; hence, the result remains independent of the cutoff scale when we fix the energy density. As for the finite part of the energy loss, we have
\be
-{\Big(\frac{dE}{dx}\Big)}^{\mathrm{norm2}}=\frac{g^2m_D^2}{6\pi}\left[{\cal A}_0(v,g)+\frac{\xi}{3}(1-3\cos^2\alpha){\cal A}_1(v,g)\right]\,.
\ee
As one can see, unlike both the unnormalized result in Eq.~(\ref{reun}) and the fixed-number-density result in Eq.~(\ref{reno}), the above result contains no monopole contribution in anisotropy correction. The corresponding correction given by
\be\label{dno2}
-{\Big(\frac{dE}{dx}\Big)}^{\mathrm{norm2}}_{(1)}=\xi \frac{g^2m_D^2}{18\pi}(1-3\cos^2\alpha){\cal A}_1(v,g)\, ,
\ee
is entirely determined by the quadrupole angular structure.

Obviously, regardless of whether and how the anisotropic distribution function is normalized, the quadrupole angular dependence of the anisotropic correction is the same in all the cases considered in this work. By contrast, monopole contribution to the anisotropy correction is sensitive to the prescription used to match the anisotropic and isotropic media. In general, we expect that the collisional energy loss varies monotonically as the incident angle increases from $0$ to $\pi/2$. Whether the energy loss is a monotonically increasing or decreasing function of $\alpha$ depends on both the incident velocity and the coupling constant, as illustrated in Fig.~\ref{reg1}. Moreover, with the fixed energy density, a notable feature arises if the quadrupole contribution is zero, in which case the anisotropy does not modify the collisional energy loss. This occurs either for a special incident angle satisfying $\cos^2\alpha=1/3$ or for the angle-averaged energy loss.

\begin{figure*}[t]
    \begin{center}
    \includegraphics[width=0.6\linewidth]{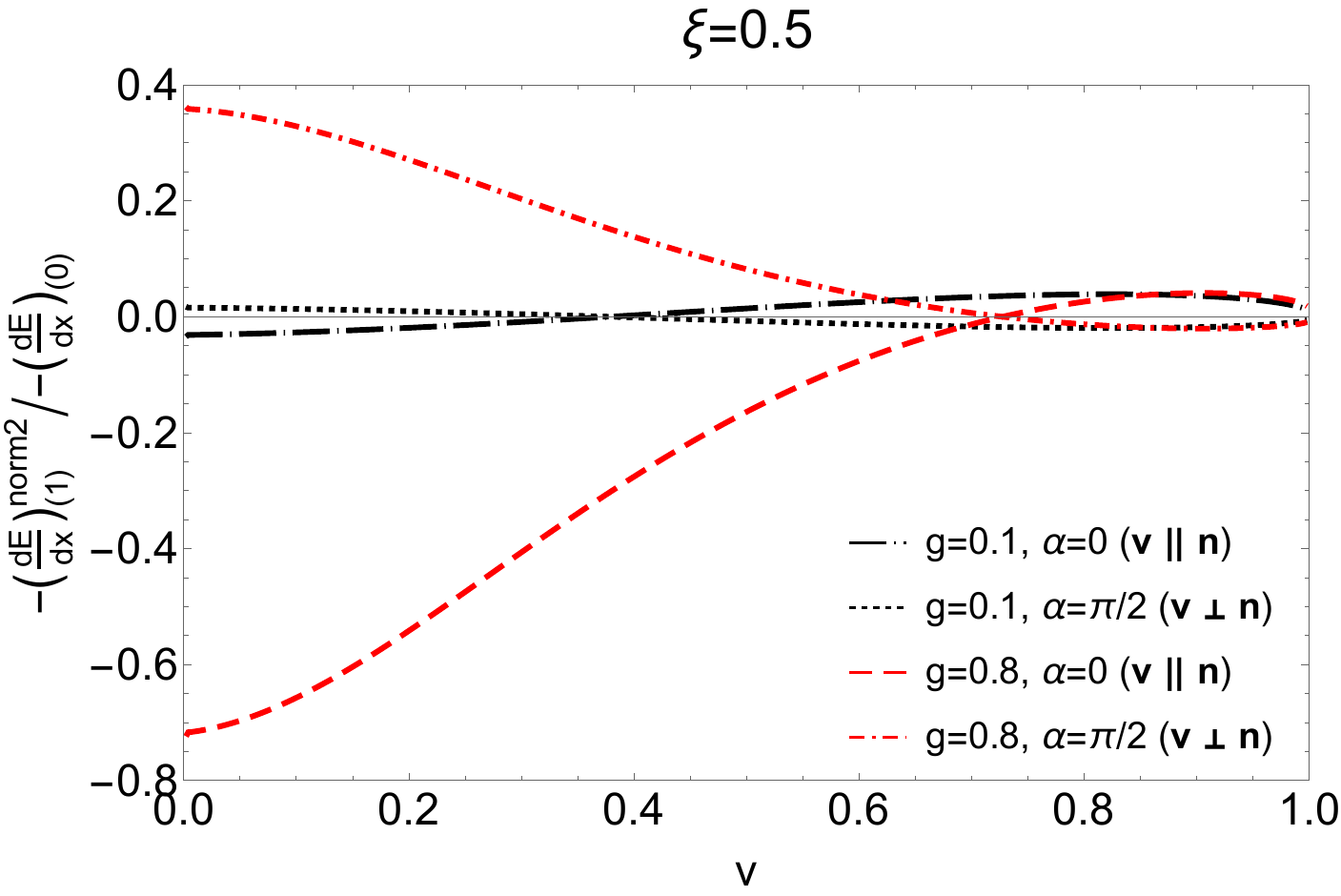}
    \end{center}
    \vspace{-5mm}
    \caption{The ratio of the anisotropy correction to the isotropic energy loss as a function of $v$ for different $g$ and $\alpha$. The results are obtained while keeping the energy density equal to that of the isotropic medium and the anisotropy parameter $\xi=0.5$.}
\label{ra3}
\end{figure*}

Eq.~(\ref{dno2}) shows that the anisotropy correction vanishes whenever $P_2(\cos\alpha)=0$ or $\mathcal{A}_1(v,g)=0$. In the angular range
$0\leq\alpha\leq\pi/2$, the first condition defines a special angle, corresponding to $\cos^2\alpha=1/3$, at which the correction
vanishes for all $v$ and $g$. For every other angle, the sign-change boundary is the same curve $\mathcal{A}_1(v,g)=0$ shown in Fig.~\ref{reg1}. For $\cos^2\alpha >1/3$, i.e., $0\le\alpha < 55^\circ$, the correction is negative above the boundary and positive below it. For $55^\circ <\alpha \le 90^\circ$,
the signs are reversed. The relative anisotropy correction shown in Fig.~\ref{ra3} further supports the results discussed above. As can be seen, for a fixed coupling constant larger than $\sim 0.03$, the anisotropy correction changes sign at the same incident velocity for arbitrary $\alpha$ (although only two specific angles are presented in the figure). This incident velocity also serves as a turning point: above or below it, the collisional energy loss is maximal for a heavy quark propagating parallel or perpendicular to the anisotropy direction, respectively. In the weak-coupling limit, we can show that
\be
\frac{-\big(dE/dx\big)^{\rm norm2}_{(1)}}{-\big(dE/dx\big)_{(0)}}\xrightarrow{g\ll 1}\frac{\xi}{3} (1-3\cos^2\alpha)\frac{{\cal J}_1(v)}{{\cal J}_0(v)}\, ,
\ee
which lies in the range of $(-\xi/15,2\xi/15)$. Since ${\cal J}_0(v)>0$ and ${\cal J}_1(v)<0$, it therefore follows that the anisotropy enhances the collisional energy loss for incident angle $0\le\alpha < 55^\circ$, whereas it suppresses the energy loss for $55^\circ<\alpha \le 90^\circ$. At the special angle $\alpha\approx 55^\circ$, the correction vanishes.

Finally, we note that unlike the equilibrium case, specifying a temperature-like scale $\lambda$ doesn't uniquely determine other thermodynamic properties of the system, such as the number density or energy density, since both will change with the anisotropy parameter $\xi$. Consequently, the collisional energy loss in an anisotropic QGP can be studied by holding either the number density or the energy density constant. In the fixed-number-density prescription discussed in the previous subsection, this is achieved by introducing the normalization factor $\sqrt{1+\xi}$ into the distribution function. However, this prescription does not preserve the energy density. On the other hand, the energy density can be held fixed at its thermal-equilibrium value by adjusting the temperature-like scale $\lambda$. These two normalization procedures lead to different modifications of the energy loss, even though both preserve the number density to linear order in $\xi$. Fixing the energy density therefore provides a complementary perspective to the fixed-number-density analysis and helps distinguish the effects associated with changes in the overall energy scale of the medium from those arising from its momentum-space anisotropy.

\section{Summary}\label{summary}

In this work, we have revisited the collisional energy loss of a heavy quark propagating through a momentum-space anisotropic quark-gluon plasma. Within the Braaten–Thoma framework, we separated the energy loss into hard and soft contributions and evaluated the leading correction in the anisotropy parameter $\xi$. Working within the weak-coupling and heavy-quark kinematic approximations specified in this paper, we derived expressions that include both logarithmic contributions and finite terms at the order considered. As in the isotropic case, the result of the hard process was fully given by analytic expressions, whereas that of the soft process was represented by a one-dimensional integral.

A central result of this work is the explicit cancellation of the artificial momentum separation scale \(q^*\) between the hard and soft contributions. Under the hierarchy \(m_D\ll q^*\ll T\), we demonstrated analytically that the logarithmic cutoff dependence also cancels in the presence of a small momentum-space anisotropy. Thus, thermal isotropy is not necessary for this cancellation at linear order in \(\xi\). The same cancellation persists when the anisotropic distribution is normalized to preserve the parton number/energy density. 

We have further shown that the angular structure of the anisotropy correction is governed by $P_2(\cos\alpha)$. This structure implies that the extrema of the energy loss occur for parallel or perpendicular propagation. Their ordering is determined by the sign of the coefficient ${\cal A}_{1}(v,g)$, which depends simultaneously on both $v$ and $g$. In the asymptotic weak-coupling limit, a heavy quark loses more energy along the anisotropy direction than perpendicular to it. A numerical exploration of the resulting expressions at moderately small couplings also reveals a reversal of this ordering in part of the low-velocity region as shown in Fig.~\ref{reg1}. Consequently, a preferred direction for maximal energy loss cannot be specified independently of the velocity and the coupling.

The sign of the anisotropy correction, which determines whether the energy loss is enhanced or suppressed relative to the isotropic result, depends on how the anisotropic and isotropic media are matched. In the asymptotic weak-coupling limit, the anisotropy correction suppresses the collisional energy loss at all propagation angles when the distribution is deformed at a common momentum scale without density normalization, whereas the correction enhances the energy loss if the parton number density is fixed by using a multiplicative factor $\sqrt{1+\xi}$. When the energy density is fixed, the monopole correction vanishes and the remaining quadrupole term enhances the energy loss for $\cos^2\alpha > 1/3$ and suppresses it for $\cos^2\alpha < 1/3$. At finite but weak coupling, determining the sign of the anisotropy correction becomes more involved, as it depends jointly on the heavy-quark speed, the coupling constant, and the propagation angle, as shown in Figs. 3 and 5 for the unnormalized result and the fixed-number-density result, respectively. To isolate the directional effect at linear order in $\xi$, a more direct way of comparing the isotropic and anisotropic media is to match the energy density. In this case, the correction remains purely quadrupolar, and all incident directions with $P_2(\cos\alpha)\neq 0$ therefore share a single sign-change boundary in the $v$–$g$ plane, as shown in Fig. 2. The anisotropy correction changes sign upon crossing this boundary curve, and the roles of parallel and perpendicular propagation as the directions of maximal and minimal energy loss interchange upon crossing the same curve. 

In closing, we note that relaxing the small-anisotropy approximation would make it possible to study the matching procedure and the angular structure for stronger deformations. In addition, a treatment of the time-dependent medium, including the evolution of anisotropy and unstable collective modes, would help connect the present local calculation to an expanding QGP. Ultimately, incorporating the collisional contribution together with radiative processes into heavy-quark transport calculations would allow the influence of momentum-space anisotropy to be investigated on heavy-flavor observables.

\section*{Acknowledgments}
The work of Y.G. is supported by the NSFC of China under Project No. 12465022; of E.W. by the NSFC of China under Project No. 12635009. 

\appendix
\section{Calculation of the hard contribution to the collisional energy loss}\label{appa}

In this appendix, we present some details of the calculation of $-\big(dE/dx\big)_{{\rm hard}(1)}$. For the $t$-channel contribution\footnote{We consider the quark-quark scattering as an example, the quark-gluon scattering can be computed very similarly.}, using Eq.~(\ref{defhard}) and the explicit form of the squared matrix element, we obtain the following expression under the approximation that $T\ll M$, $T/E\ll v$ and $E\ll M^2/T$,
\ba
-\bigg(\frac{dE}{dx}\bigg)^{Qq}_{{\rm hard}(1)}&=&-\frac{\xi}{\lambda v}\frac{ 4\pi}{3}g^4 N_f \int \frac{d^3 {\bf k}}{(2\pi)^3 }  \frac{({\bf n}\cdot{\bf k})^2}{k^2}(1- n_F(k))n_F(k)\int \frac{d^3 {\bf {k}}'}{(2\pi)^3 } \frac{\omega}{k^\prime} \,\nonumber \\
&\times& \frac{1}{(\omega^2-q^2)^2}\Big[2(k-{\bf{v}}\cdot {\bf {k}})^2+\frac{1-v^2}{2}(\omega^2-q^2)\Big]\delta(\omega-{\bf v}\cdot{\bf q})\, .
\label{hard1qq}
\ea
To average over the solid angles, according to Eq.~(\ref{id}), two general forms that we need to consider are given by
\be\label{id2}
\int_{\Omega_{\bf n}}({\bf n}\cdot{\bf k})^2\delta\left(\frac{{\bf v}\cdot {\bf n}}{v}-\cos \alpha\right)=\frac{\pi}{2}\left[k^2(1-\cos 2\alpha)+\frac{({\bf v}\cdot {\bf k})^2}{v^2}(1+3\cos 2\alpha)\right]\, ,
\ee
and
\be
\int_{\Omega_{\bf v}}({\bf v}\cdot{\bf k})^s\delta(\omega-{\bf v}\cdot{\bf q})=\frac{2\pi}{vq}{\cal I}^{(s)}
\ee
\begin{equation}\la{id3}
{\cal I}^{(s)}=\left\{
\begin{aligned}
&1 \quad &{\rm for}\quad s=0\\
&\frac{\omega }{q^2}({\bf{k}}\cdot{\bf{q}})  &\, {\rm for}\quad s=1\\
&\frac{1}{2 q^4}\Big[(v^2q^2-\omega^2) k^2 q^2+(3\omega^2-v^2q^2)(\mathbf{k}\cdot\mathbf{q})^2\Big]
&\,{\rm for}\quad s=2 \\
&\frac{\omega(\mathbf{k}\cdot\mathbf{q})}{2 q^6}
\Big[ 3(v^2q^2-\omega^2)k^2 q^2 +(5\omega^2-3v^2q^2)(\mathbf{k}\cdot\mathbf{q})^2 \Big]
&\,{\rm for}\quad s=3\\
&\frac{1}{8 q^8}
\Big[3
\Big(
(v^2q^2-\omega^2)\big(k^2q^2-(\mathbf{k}\cdot\mathbf{q})^2\big) +4\omega^2(\mathbf{k}\cdot\mathbf{q})^2\Big)^2
-40\omega^4(\mathbf{k}\cdot\mathbf{q})^4
\Big]
&\,{\rm for}\quad s=4
\end{aligned}
\right.\,.
\end{equation}
As a result, we can rewrite Eq.~(\ref{hard1qq}) in terms of ${\cal I}^{(s)}$ as
\ba
-\left(\frac{dE}{dx}\right)^{Qq}_{{\rm hard}(1)}
&=&-\frac{\xi}{\lambda v}\frac{4\pi}{3}g^4N_f\int \frac{d^3\mathbf k}{(2\pi)^3}
\int \frac{d^3\mathbf k'}{(2\pi)^3}(1-n_F(k))n_F(k)\frac{\omega}{k'}\frac{1}{(\omega^2-q^2)^2}\nonumber
\\&\times&\frac{1}{4vq}\Bigg\{(1-\cos2\alpha)\left[\left(k^2+\frac{1-v^2}{4}(\omega^2-q^2)\right){\cal I}^{(0)}-2k\, {\cal I}^{(1)}+{\cal I}^{(2)}\right]
\nonumber \\
&+&\frac{1+3\cos2\alpha}{v^2 k^2}\left[\left(k^2+\frac{1-v^2}{4}(\omega^2-q^2)\right){\cal I}^{(2)}-2k\, {\cal I}^{(3)}+{\cal I}^{(4)}\right]\Bigg\}\,.
\ea
Clearly, the integrand now depends on $k$, $k^\prime$ and $\cos \theta_{k^\prime}$, where $\theta_{k^\prime}$ is the angle between ${\bf k}$ and ${\bf k}^\prime$, i.e., the polar angle of ${\bf k}^\prime$ when ${\bf k}$ is along the $z$-axis. The remaining steps are closely analogous to those in the isotropic case. More specifically, we rewrite this integrand as a function of $k$, $q$ and $\omega$ with $\cos\theta_{k^\prime}=(k^2+(\omega+k)^2-q^2)/(2k(\omega+k))$, and also change the integration variables according to Eq.~(\ref{changeva}). One can then analytically carry out the successive integrals over $\omega$, $q$, and $k$, leading to the result as given in Eq.~(\ref{qqtEhard}). 

For the contributions from the $s$- and $u$-channels in quark-gluon scattering, one can also follow the same strategy as for the $t$-channel calculation by performing the angular average. However, since the scalar product ${\bf v}\cdot{\bf k}$ appears in the denominator, the angular average leads to a very complicated expression, which makes the subsequent integrals over $k$, $q$, $\omega$ rather intractable. Therefore, for the $s$- and $u$-channels,  it turns out to be simpler to proceed without considering the angular average. With the explicit form of the squared matrix element, we have
\ba
-\left(\frac{dE}{dx}\right)^{Qg(s,u)}_{{\rm hard}(1)}&=&-\xi\frac { \pi g ^ {4}}{ vT}\int\frac{d^3\mathbf k}{(2\pi)^3}
\int \frac{d^3\mathbf k'}{(2\pi)^3} \frac {(\mathbf {k} \cdot {\mathbf {n}}) ^ {2}}{k ^ {2} k ^ {\prime}} [ 1 + n _ {B} (k) ] n _ {B} (k) \frac {(1 - v ^ {2}) ^ {2}}{(k - \mathbf {k} \cdot \mathbf {v}) ^ {2}}\nonumber \\ &\times& (k ^ {\prime} - k) \delta (k ^ {\prime} - k - \mathbf {v} \cdot (\mathbf {k} ^ {\prime} - \mathbf {k})). 
\ea
Since we don't change the integration variables in the following evaluation, the integrand in the above equation is expressed as a function of ${\bf k}$ and ${\bf k}^\prime$. By choosing the polar axis to be along the direction of ${\bf v}$, we can first perform the integral over $k^\prime$ using the delta function. As a result, the radial and angular integrals decouple and we arrive at
\ba
-\left(\frac{dE}{dx}\right)^{Qg(s,u)}_{{\rm hard}(1)} &=& -\,\xi\,\frac{\pi g^4 (1-v^2)^2}{v T(2\pi)^6}\left[\int_0^\infty dk\,k^2[1+n_B(k)]n_B(k)\right] {\rm Int}_{\Omega}\nonumber \\
&=&-\,\xi\,\frac{ g^4 T^2 (1-v^2)^2}{192\pi^3 v}\, {\rm Int}_{\Omega}\, ,
\ea
where
\be
{\rm Int}_{\Omega} = \int d \Omega_{\bf k} d \Omega_{{\bf k}^ {\prime}} \left(\cos \alpha\, \mu+\sqrt{1-\mu^2}\sin \alpha \cos\phi_k\right)^2 \frac {(1 - v \mu) - (1 - v \mu^ {\prime})}{(1 - v \mu) (1 - v \mu^ {\prime}) ^ {3}}\, .
\ee
In the above equation, we use the shorthand notations $\mu$ and $\mu^\prime$ to denote the cosines of the polar angles of ${\bf k}$ and ${\bf k}^\prime$, respectively.

Integrating over the azimuthal angles, we find that ${\rm Int}_{\Omega}$ becomes
\ba
{\rm Int}_{\Omega} &=& 2 \pi^ {2} \int_{-1}^{1}d\mu\int_{-1}^{1}d\mu' \left[\sin^2\alpha + (3 \cos^2\alpha - 1)\mu^2 \right]\frac{(1-v\mu)-(1-v\mu')}{(1-v\mu)(1-v\mu')^3}\nonumber \\
&=&-\frac{4\pi^2}{3}\left[B_1(v)B_2(v)-2B_3(v)
+(1-3\cos^2\alpha)B_2(v)\left(\frac{v^2-3}{2v^2}B_1(v)+\frac{3}{v^2}\right)\right]\,.
\ea
with 
\be
B_n(v)=\int_{-1}^{1}\frac{d x}{(1-v x)^n}\,.
\ee
Using the analytical forms of $B_n(v)$ for $n=1,2,3$, it is easy to show the result as obtained in Eq.~(\ref{qgsuEhard}).

\bibliographystyle{apsrev4-1}
\bibliography{elaniso}
\end{document}